\documentclass[12pt]{revtex4}
\usepackage{graphicx}
\usepackage{color}
\usepackage{amssymb}
\usepackage{amsmath}

\newcommand{\be}{\begin{equation}}
\newcommand{\ee}{\end{equation}}
\newcommand{\ben}{\begin{eqnarray}}
\newcommand{\een}{\end{eqnarray}}
\newcommand{\bes}{\begin{subequations}}
\newcommand{\ees}{\end{subequations}}
\newcommand{\bb}{\bibitem}

\newcommand{\wt}{\widetilde}

\begin{document}
\title[Deformation procedure for scalar fields in cosmology]{Deformation procedure for scalar fields in cosmology}
\author{D. Bazeia$^1$, L. Losano$^1$ and A. B. Pavan$^2$}\email{alan@unifei.edu.br}
\address{$^1$ Departamento de F\'\i sica, Universidade Federal da Para\'\i ba, 58051-970 Jo\~ao Pessoa PB, Brazil}
\address{$^2$ Instituto de F\'isica e Qu\'imica, Universidade Federal de Itajub\'a, CEP 37500-903, Itajub\'a, MG, Brazil}

\begin{abstract}
This work offers an extension of the deformation procedure introduced in field theory to the case of standard cosmology in the presence of real scalar field in flat space-time. The procedure is shown to work for different cosmological models, which give rise to several different cosmic scenarios, evolving under the presence of first-order differential equations which solve the corresponding equations of motion very appropriately. Additionally, it provides an interesting approach to introduce, in the scalar field potential, a non-linear interaction of the scalar field. 
\end{abstract}
\maketitle

\section{Introduction}

The recent discovery of cosmic acceleration \cite{a1,a2,a3} has brought to the stage one the main mysteries of modern cosmology.
Named dark energy, it amounts to almost $68\%$ of the total energy of the universe to cause the present cosmic acceleration.
According to the recent understanding \cite{prr,pr,tc,li,cst}, one of the main fuel of the cosmic acceleration is quintessence
\cite{q1,q2,q3,q4}, which in standard Friedmann-Lemaitre-Robertson-Walker (FLRW) cosmology may be described by real scalar fields.

The presence of acceleration has raised a great deal of interest in cosmology driven by scalar fields, welcoming the construction
of new cosmological models. Among the several possibilities, which are reviewed in Refs.~{\cite{prr,pr,tc,li,cst}}, we are mainly interested in FLRW cosmology driven by real scalar 
field with standard dynamics. In particular, the specific aim of the present work is inspired in investigations concerning the accelerated expansion of the Universe, as considered in \cite{pr,tc,li,cst,padma,new0,FT,Nori,new,new1} and in references therein.

Before going into the main issue, however, let us recall another line of investigation which develops alternative methodology. In the recent work \cite{bglm} some of us have 
turned attention to FLRW models described by real scalar fields, focusing on inspecting Einstein's equation and the equation of motion for the scalar field in a very direct way, 
considering models described by real scalar field in generic space-time, displaying the usual spherical, flat or hyperbolic spatial profile. The pioneer investigations show how to 
obtain first-order differential equations which solve the corresponding equations of motion, for cosmology in general spherical, flat or hyperbolic geometry. The presence of 
first-order equations was then extended to brane in \cite{abl}, and this has also been explored in Refs.~{\cite{st,blr,st2}} in several contexts, including the case of brane with 
a single extra dimension, which has also appeared in the former investigations \cite{b1,b2,b3,b4,b5,b6,b7}. A cosmological model with interacting dark matter and dark energy was developed using first order formalism in \cite{alan}.
Evidently, the unveiling of first-order equations ease the process of 
solving specific models, finding analytical solutions which can be used to fully understand the related cosmic evolution.

The first-order formalism offers an interesting way to investigate FLRW models in standard cosmology. In the present
investigation our main motivation is to use the first-order formalism to extend the deformation procedure of Ref.~\cite{blm} to modern
cosmology. We believe that the possibility of deforming the cosmic evolution of some standard model will certainly lead to new
possibilities of investigation, contributing to improve the present view of the subject. 
Indeed, in some recent works \cite{def1,def2} one has shown how to extend the deformation procedure of Ref.~{\cite{blm}} to several other scenarios, including the Randall-Sundrum 
braneworld scenario \cite{rs} described by warped geometry with a single extra dimension of infinity extend, and these results may help us to further enlarge the deformation 
procedure to the case of braneworld with internal structure, as investigated for instance in Refs.~{\cite{bi1,bi2}}. Another interesting application of the deformation procedure has appeared more recently in \cite{Blinov}, concentrating on the asymptotic properties of kinks.

The key motivation for the present work has appeared after the work \cite{padma}, which deals with FLRW cosmology driven by real scalar field.
The scalar field model is driven by some potential $V=V(\phi).$ In this case, if the scale factor $a(t)$ is known, the potential $V(t)$ and the field $\phi(t)$ are also known 
\cite{padma}. Thus, if one inverts $t=t(\phi)$ we can write $V=V(\phi).$ The lesson to learn here is that the cosmic evolution determines the potential. This is a very nice 
result, and we can use it together with the deformation procedure of Ref.~\cite{blm}, which introduces a direct approach to relate different models, described by different 
potentials. There we learn how to get from $V(\phi)$ to another potential, the deformed potential $\wt V(\wt\phi).$ Thus, if we relate $V(\phi)$ to $a(t)$, then we can also relate 
$\wt V(\wt\phi)$ to ${\wt a}(t),$ and so we can relate $a(t)$ to ${\wt a}(t).$ This is the deformation procedure, now extended to the cosmic evolution. As we show below, it works 
nicely for standard FLRW cosmology in flat space-time, for models described by first-order differential equations.

To make the deformation procedure effective, for simplicity we deal with flat geometry, that is, we consider the line element
\begin{equation}
 ds^2=dt^2-a^2(t)[dr^2+r^2d\Omega^2]
\end{equation}
where $a(t)$ is the scale factor, and we also use $H={\dot a}/a$ to describe the Hubble parameter. This is the standard
scenario, and now we include scalar field model with a single real scalar field. In this case, the Einstein-Hilbert action is given by
\begin{equation}
\label{model}
S=\int\,d^4x\;{\sqrt{-g}\;\left(-\frac14\,R+{\cal L(\phi,\partial_\mu\phi)}\right)}
\end{equation}
where $\phi$ describes the scalar field and we are using ${4\pi G}=1.$ In general, the energy-momentum tensor is given by
$T^\mu_{\;\;\nu}=diag(\rho,-p,-p,-p),$ where $\rho$ and $p$ represent energy density and pressure. We use Einstein's equation to get
\begin{eqnarray}
\label{fe}
\frac{\ddot a}{a}&=&-\frac13\,(\rho+3p),
\\
\label{fe1}
H^2&=&\frac23\,\rho.
\end{eqnarray}
The equation of motion for the scalar field depend on ${\cal L}(\phi,\partial_\mu\phi),$ which has the standard form
\begin{equation}
\label{sm}
{\cal L}=\frac12\partial_\mu\phi\partial^\mu\phi-V(\phi)
\end{equation}
The energy density and pressure that account for the scalar field model are given by
\begin{equation}
\rho=\frac12\dot\phi^2+V,\;\;\;\;\;p=\frac12\dot\phi^2-V
\end{equation}
and the equation of motion for the scalar field is given by
\begin{equation}\label{em1}
\ddot\phi+3H\dot\phi+\frac{dV}{d\phi}=0
\end{equation}
We use the energy density and pressure to rewrite Eqs.~(\ref{fe}-\ref{fe1}) in the form
\begin{eqnarray}
\label{em2}
{\dot H}=-\dot\phi^2,
\\
\label{em3}
H^2=\frac13{\dot\phi}^2+\frac23V
\end{eqnarray}
As one knows, the above Eqs.~(\ref{em1}-\ref{em3}) constitute the equations that we have to solve in the case of FLRW cosmology driven by scalar field with standard dynamics in flat space-time.

In this standard scenario in cosmology, the deformation procedure that we develop below may have connections with other interesting investigations. Some direct issues concern 
Refs.~{\cite{barrow95,sahni}}: in the first case, in Ref.~{\cite{barrow95}} one investigates the generation of new solutions from existing solutions, taking advantage of the 
presence of invariance of Einstein's equations in FLRW models described by a single real scalar field in flat space-time; in the second case, in Ref.~{\cite{sahni}} one deals with cosmological reconstruction of scalar field potential and dark energy. These investigations introduce new motivations and enlarge our interest in the subject.

\section{First-order equations}
\label{s2}

Before going to the subject we want to explore in the present work, let us firstly review the main
steps of the first-order formalism put forward in Refs.~\cite{bglm,blr}. In FLRW cosmology, it is the standard view that the scale factor, the scalar field and the Hubble parameter 
are all functions of the time evolution, that is, $a=a(t),$ $\phi=\phi(t),$ and $H=H(t).$ Also, from Einstein's equation we need to see the potential as a function of time too. 
However, from the equation of motion for the scalar field we have $V=V(\phi),$ which shows the potential as a function of the scalar field. Thus, to make these two views 
equivalent we can also think of the Hubble's parameter as a function of the scalar field. This is the key point, and we make it very efficient with the introduction of a new 
function -- $W=W(\phi)$ -- from which we can now understand that the Hubble parameter depends on time as a function of $W[\phi(t)].$

\subsection{The formalism}

To make the above reasoning formal, we introduce $W=W(\phi)$ such that the mathematical expression which follows can be written as
\be\label{be0} 
H=W(\phi).
\ee
This is a first-order differential equation for the scale factor. It can be integrated leading to the scale factor $a(\phi)$ as function of the scalar field,
\be
\label{sfactor}
a(\phi)= \exp\left(-\int W/W_{\phi} \ d\phi+C\right)
\ee
where $W_\phi$ stands for $dW/d\phi$. The Eq. \eqref{be0} also allows us to obtain another equation, involving the scalar field, in the form
\be\label{bek}
\dot\phi=-W_\phi.
\ee
These Eqs.~(\ref{be0}) and (\ref{bek}) imply that the potential has to have the form
\be\label{pot}
V=\frac32 W^2-\frac12W_{\phi}^2.
\ee
This procedure in flat space-time is similar to the Hamilton-Jacobi approach of Ref.~\cite{ll}. If instead of the deceleration parameter,
we define ${\bar q}={\ddot a}a/{\dot a}^2$ as the acceleration parameter, we get ${\bar q}=1+{\dot H}/H^2.$ For the above model, the acceleration parameter has the form
\be \label{qe}
\bar q=1-\frac{W_\phi^2}{W^2}.
\ee
The energy density and pressure are given by, respectively
\be
\rho(t)=\frac32 W^2,\;\;\;\;\;\;p(t)=W_\phi^2-\frac32W^2,
\ee
and now we can write the equation of state $\omega=p/\rho$ in the form
\be\label{eqstate}
\omega=\frac23\frac{W_{\phi}^2}{W^2}-1.
\ee
It is not constant, and from Eqs.~(\ref{qe}) and (\ref{eqstate}) we see that for an accelerated expansion we must have $\omega<-1/3$. In order to show the worth and extension of the first order formalism  we illustrate the above results with some examples.
\subsection{First example}

Firstly, we take
\be
\label{W1} W=A\phi^{n}+B
\ee
where $A$ and $B$ are real parameters. This leads to models defined by the potentials:
\be
V(\phi)=\frac{3}{2}\left[A^{2}\phi^{4}+2AB\phi^{2}+B^2\right]-2A^{2}\phi^2
\ee
for $n=2,$ and
\be 
V(\phi)=\frac{3}{2}\left[A^{2}\phi^{2n}+2AB\phi^{n}+B^2\right]-\frac{A^{2}n^{2}\phi^{2n-2}}{2}
\ee
for $n\neq2.$ For these models, the set of Eqs.~(\ref{em1}-\ref{em3}) are solved by
\begin{eqnarray}
\label{edc}
\phi(t)&=&e^{-2At}, \;\;\;\; for\ \ n=2\\
\label{edc1}
\nonumber\\
\phi(t)&=&\left[An(n-2)t\right]^{\frac{1}{2-n}}, \;\;\;\; for\ \ n\neq2 \label{edc2}
\end{eqnarray}
and the Hubble parameters
\ben
H(t)&=&Ae^{-4At}+B,\;\;\;\; for\ \ n=2\\
\label{Hn2}
\nonumber\\
H(t)&=&A\left[An(n-2)t\right]^{\frac{n}{2-n}}+B,\;\;\;\; for\ \ n\!\neq\!2
\label{Hn}
\een
These results allow obtaining the scale factors
\be
a(t)=\exp\left(\frac{A(2-n)[An(n-2)]^{\frac{n}{2-n}}}{2}\ t^{\frac{2}{2-n}}+Bt\right)
\label{an2}
\ee
for $n\neq2,$ and
\be
a(t)=\exp\left(-\frac{1}{4}e^{-4At}+Bt\right),\;\;\;\; \text for\ \ n=2
\label{an}
\ee
Here we need to pay attention to Eqs.(\ref{edc2}), (\ref{Hn}) and (\ref{an2}) when we choose $n\neq 2,$ since it is necessary to adjust $n$ and $A$ to make all quantities real. We see that the scale factor in Eq.~(\ref{an2}) leads to intermediate inflation, an issue set forward in Ref.~\cite{barrow90}.

\begin{figure}[h]
\centering
\includegraphics[{height=6cm,width=7cm,angle=00}]{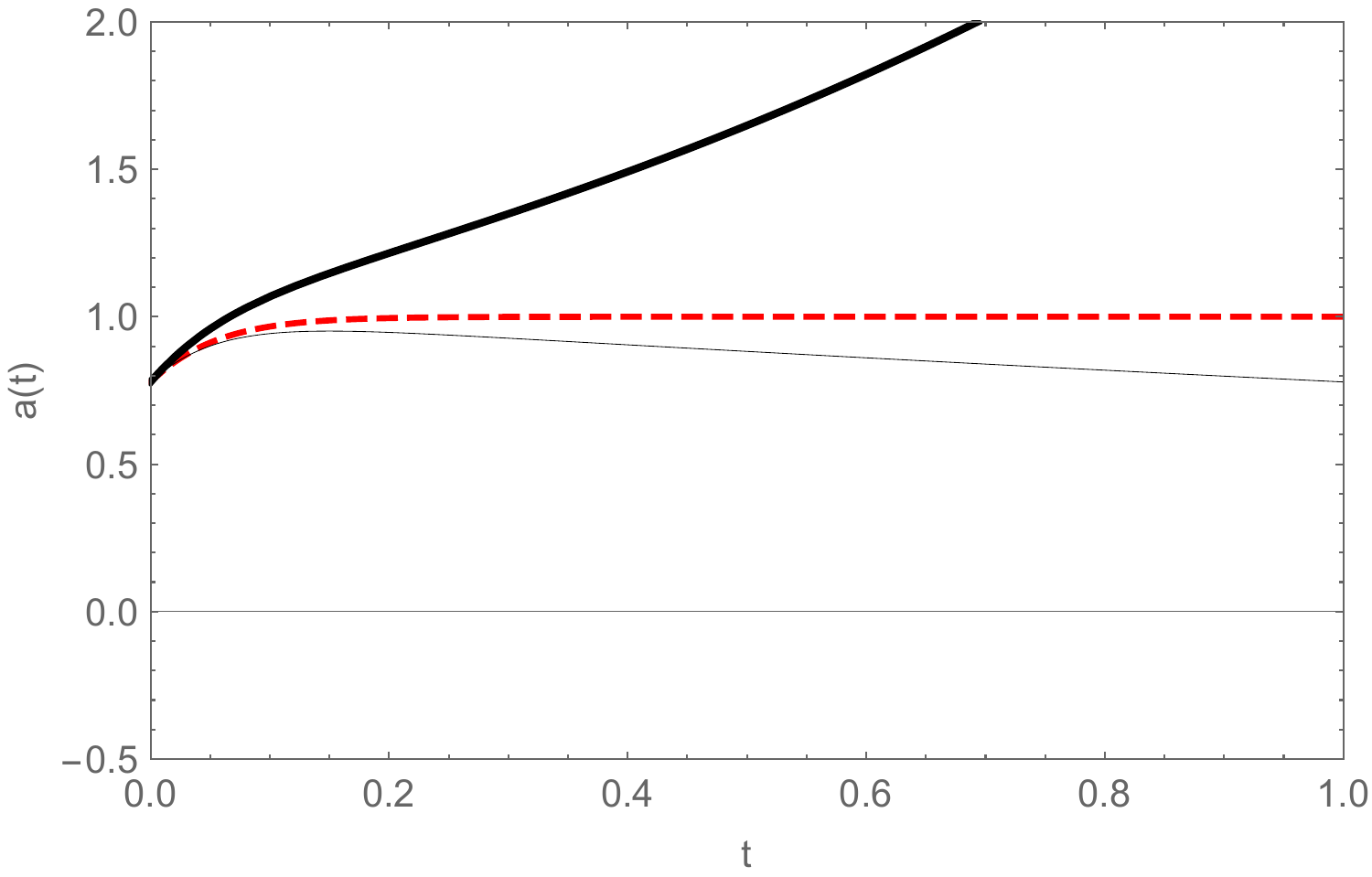}
\includegraphics[{height=6cm,width=7cm,angle=00}]{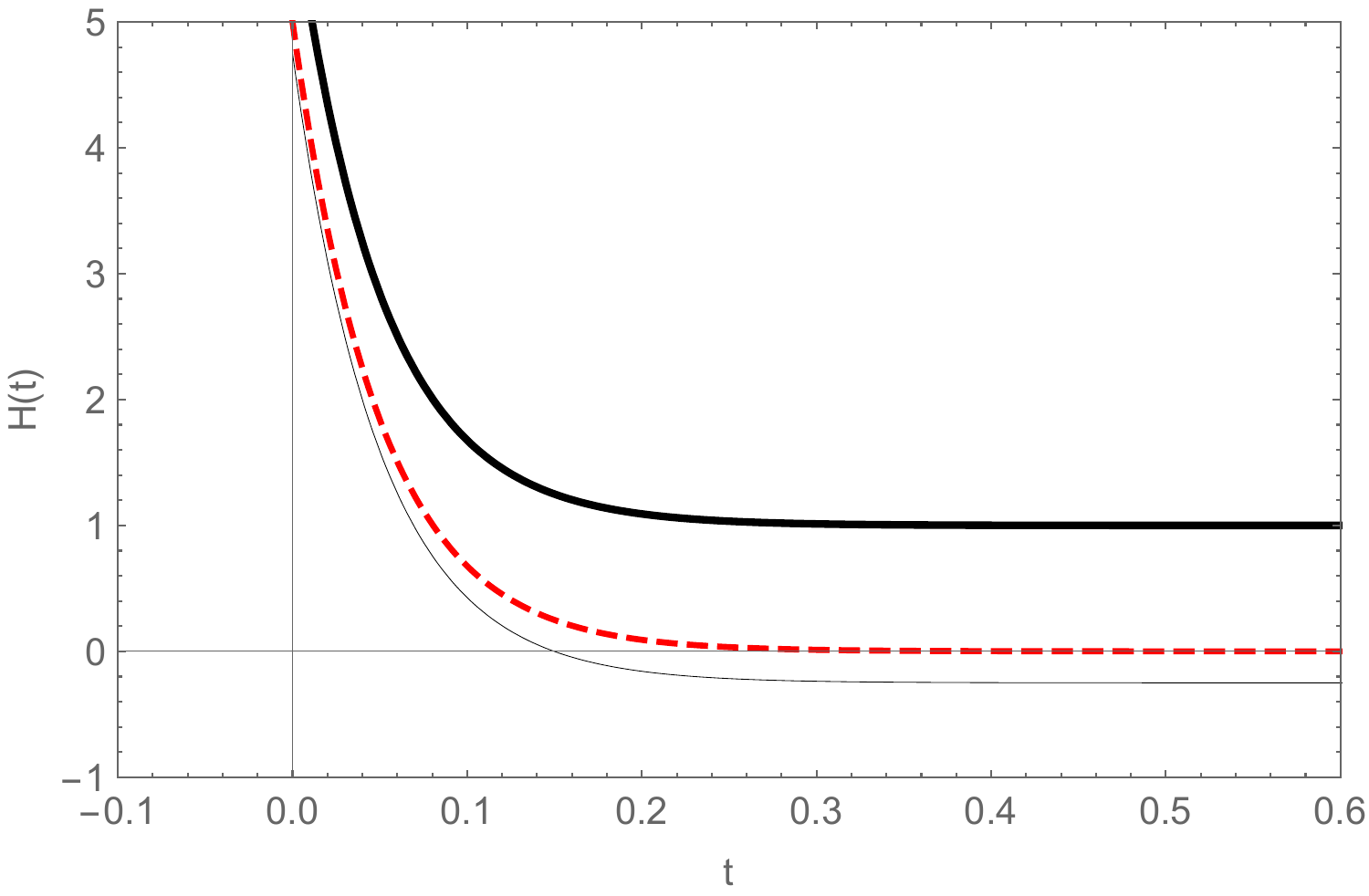}
\caption{The scale factor $a$ (left panel) and the Hubble's parameter $H$ (right panel) are plotted for $A=5$, and $B=0$ (dotted line), $B=1$ (thick line), and $B=-1/4$ (thin 
line).}
\label{fig1}
\end{figure}

In the above results, we consider the case for $n=2$. In this case, $A$ can be positive or negative. We first restrict our attention to
the case $A$ positive, and in Fig.~\ref{fig1} we display both the scale factor $a$ and Hubble's parameter $H$ for some choice of parameters.
Here we have three distinct sub-models: for $B=0$ the universe expands, asymptotically approaching the static scenario; for $B>0$
the universe expands eternally and for $B<0$ the universe turns from a expanding phase to contraction. In late times, all the models
behave according to $e^{Bt}$, that is, evolve to a de Sitter universe. For $A<0,$ the value of $B$ does not qualitatively change the
evolution, which is now similar to the former case, for $A>0$ and $B<0$.

\begin{figure}[h]
\centering
\includegraphics[{height=6cm,width=7cm,angle=00}]{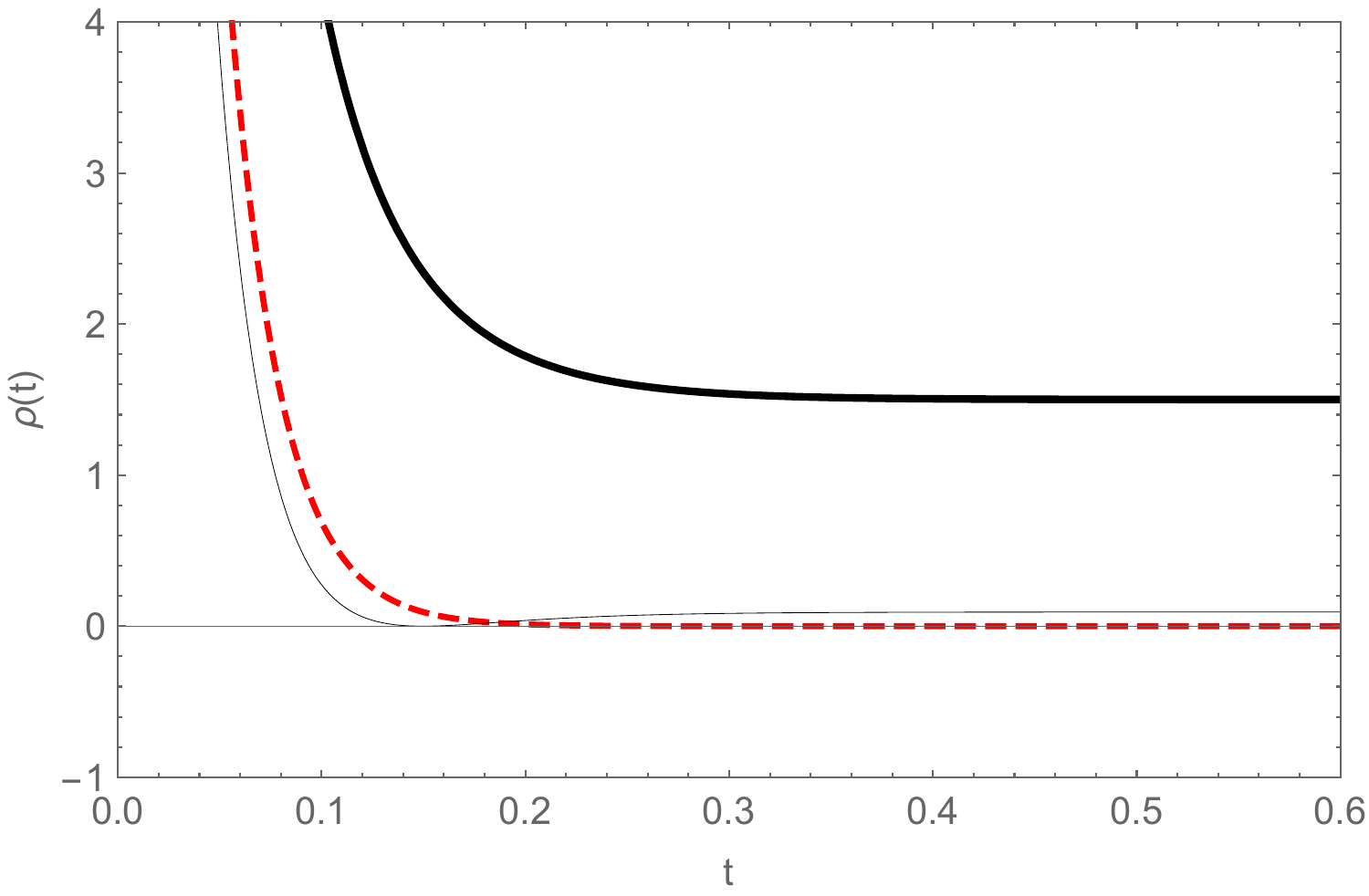}
\includegraphics[{height=6cm,width=7cm,angle=00}]{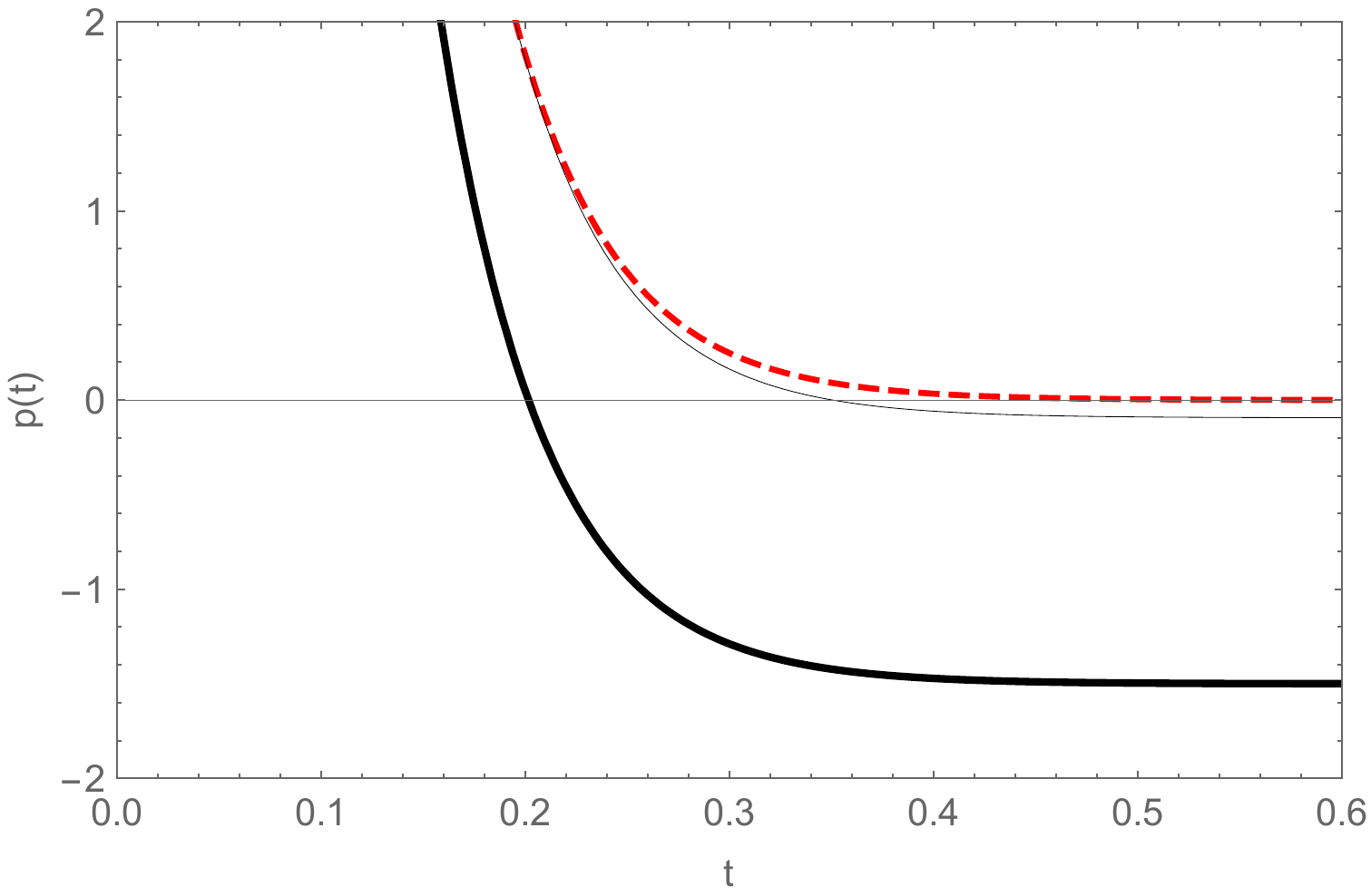}
\caption{The energy density $\rho$ (left panel) and the pressure $p$ (right panel) for $A=5$, and $B=0$ (dotted line), $B=1$ (thick line), and $B=-1/4$ (thin line).} \label{fig2}
\end{figure}

The energy density and pressure are given by
\begin{eqnarray}
\rho(t)&=&\frac{3}{2}\left[Ae^{-4At}+B\right]^{2}\\
\nonumber\\
p(t)&=&4A^{2}e^{-4At}-\frac{3}{2}\left[Ae^{-4At}+B\right]^{2}
\end{eqnarray}
In Fig. \ref{fig2} we plot the energy density $\rho$ and pressure $p$. Again, if we restrict our attention to the case $A>0,$ we can
see that: for $B=0$ the energy density and pressure are always positive, decreasing to zero asymptotically. For $B>0$ the
pressure is initially positive and it decreases asymptotically, going to a negative constant, while the energy density starts positive,
decreases to its minimum $(\rho=0)$ and increases getting asymptotically to a positive constant. This case connects two
distinct scenarios, with different matter contents. Finally, for $B<0,$ pressure and energy density behave similarly, as in the
case with $B=0.$ If $A<0,$ the energy density increases, independently of the value of $B$, becoming exponentially infinity
at later times; for the pressure, it decreases getting asymptotically to the negative infinity exponentially.


\begin{figure}[h]
\centering
\includegraphics[{height=6cm,width=7cm,angle=00}]{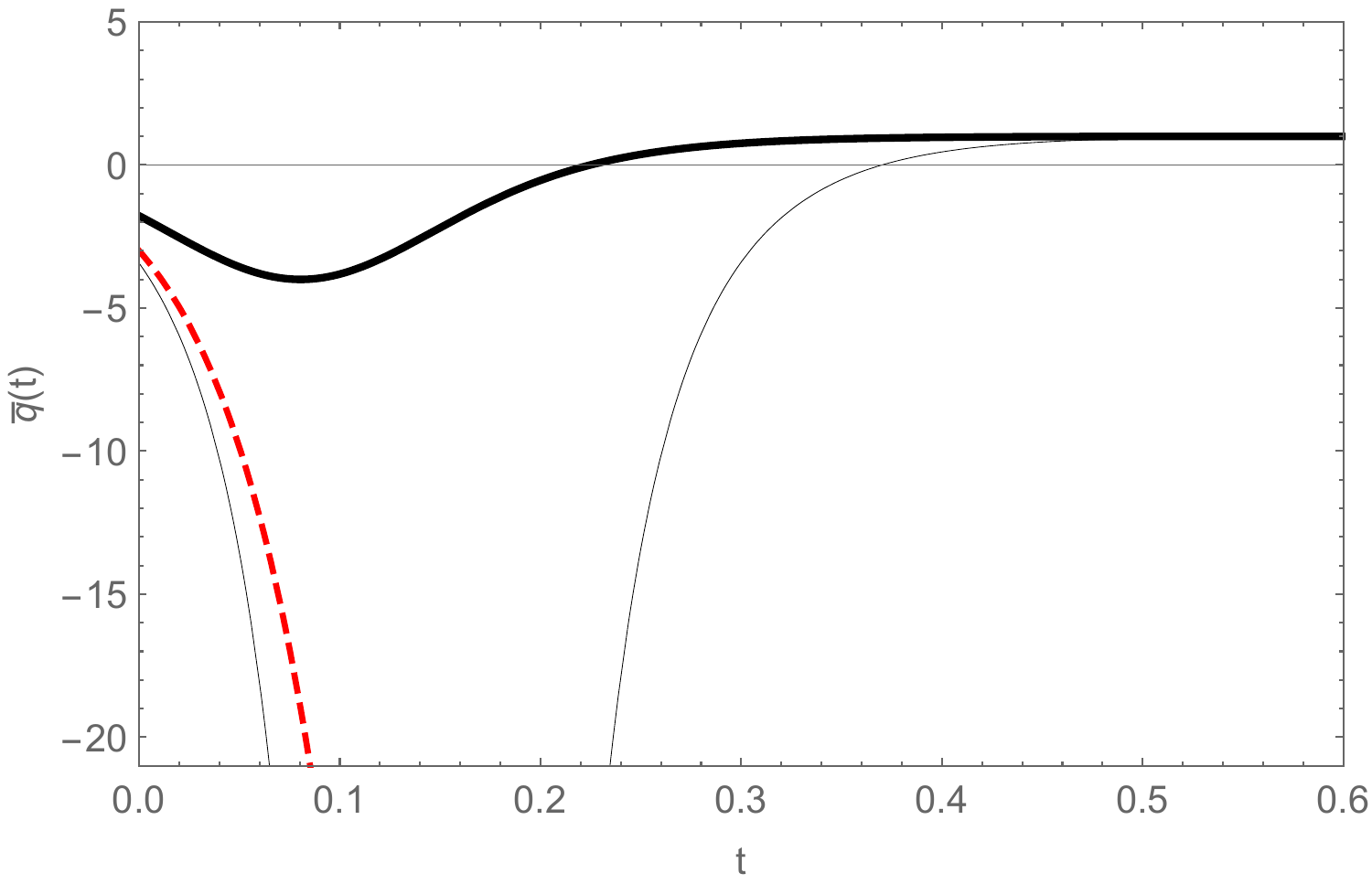}
\includegraphics[{height=6cm,width=7cm,angle=00}]{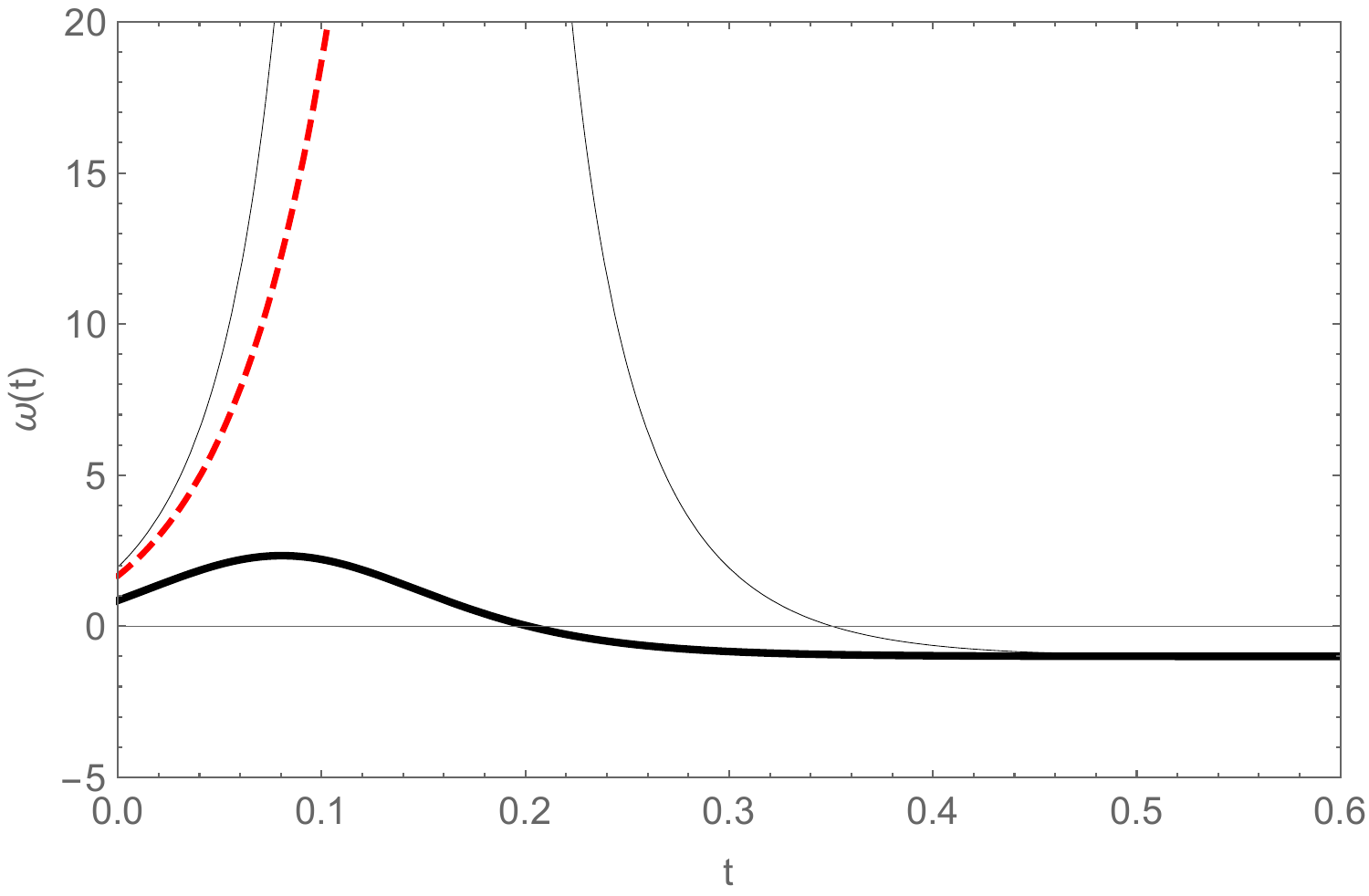}
\caption{The acceleration parameter $\bar{q}(t)$ (left panel) and the $\omega$ parameter (right panel) for $A=5$, $B=0$ (dotted line), $A=5$, $B=1$ (thick line) and $A=5$, $B=-1/4$ (thin line).}
\label{fig3}
\end{figure}

The acceleration parameter in this model is given by
\be
\bar{q}=1-\left(\frac{2Ae^{-2At}}{B+Ae^{-4At}}\right)^{2}
\ee
In Fig.~\ref{fig3} we depict the acceleration parameter $\bar{q}$ and $\omega.$ In general, for $A>0$ and with $B=0$ the cosmic evolutions are always desacelerated with $\omega>0;$ 
with $B>0$ two phases are possible: one, decelerated, followed by another phase, accelerated, with $\omega$
starting in a positive value and ending in a negative one. In the last case, with $B<0$, both $\bar{q}$ and $\omega$ may diverge, when the
energy density goes to zero and the pressure has a finite value.

\subsection{Second example}

We now introduce another example, described by
\be
\label{W2}
W=A\cosh{(B \phi)}
\ee
where $A$ and $B$ are real parameters. This choice leads to the potential
\begin{eqnarray}
V(\phi)&=&\frac{3}{2}A^2\cosh^2{(B\phi)}-\frac{1}{2}A^2B^2\sinh^2{(B\phi)}
\end{eqnarray}
Similar potential was suggested to describe quintessence and dark matter in the work \cite{sahniprd}. For the present potential, the set of Eqs.~(\ref{em1}-\ref{em3}) are 
solved by
\begin{eqnarray}
\label{phicosh}
\phi(t)&=&\frac{1}{B}\ln{\left(\tanh{\left(\frac{AB^2t}{2}\right)}\right)}
\end{eqnarray}
and the Hubble's parameter has the form
\begin{eqnarray}
H(t)&=&A\cosh{\left(\ln{\left(\tanh{\left(\frac{AB^2t}{2}\right)}\right)}\right)}
\end{eqnarray}
The scale factor becomes
\be
a(t)=\left[\tanh{\left(\frac{AB^2t}{2}\right)}^2-1\right]^{-1/B^2}\ \ \ \tanh{\left(\frac{AB^2t}{2}\right)}^{1/B^2}
\ee
We see from Eq.~(\ref{phicosh}) that $A>0.$

\begin{figure}[ht]
\centering
\includegraphics[{height=6cm,width=7cm,angle=00}]{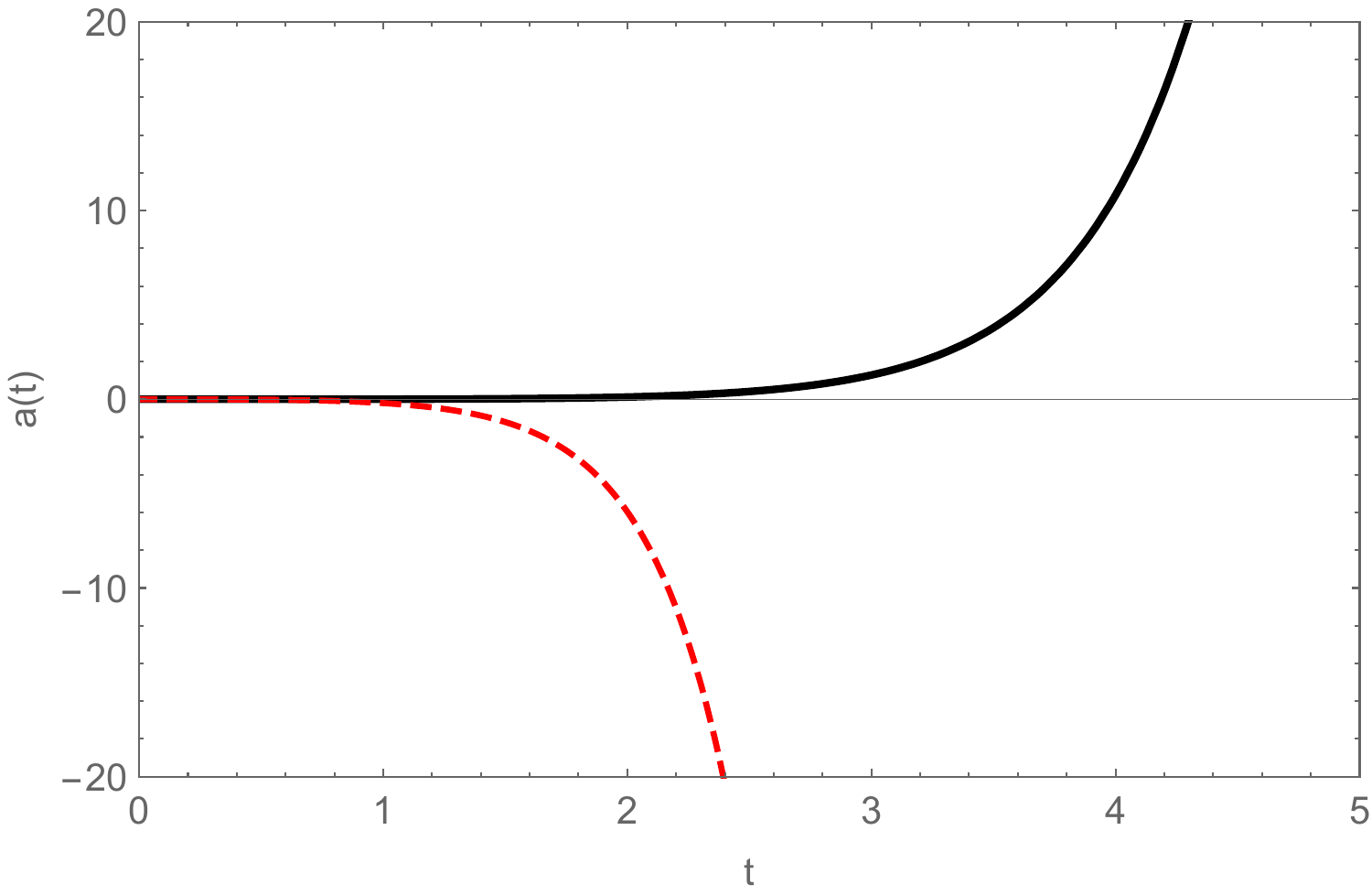}
\includegraphics[{height=6cm,width=7cm,angle=00}]{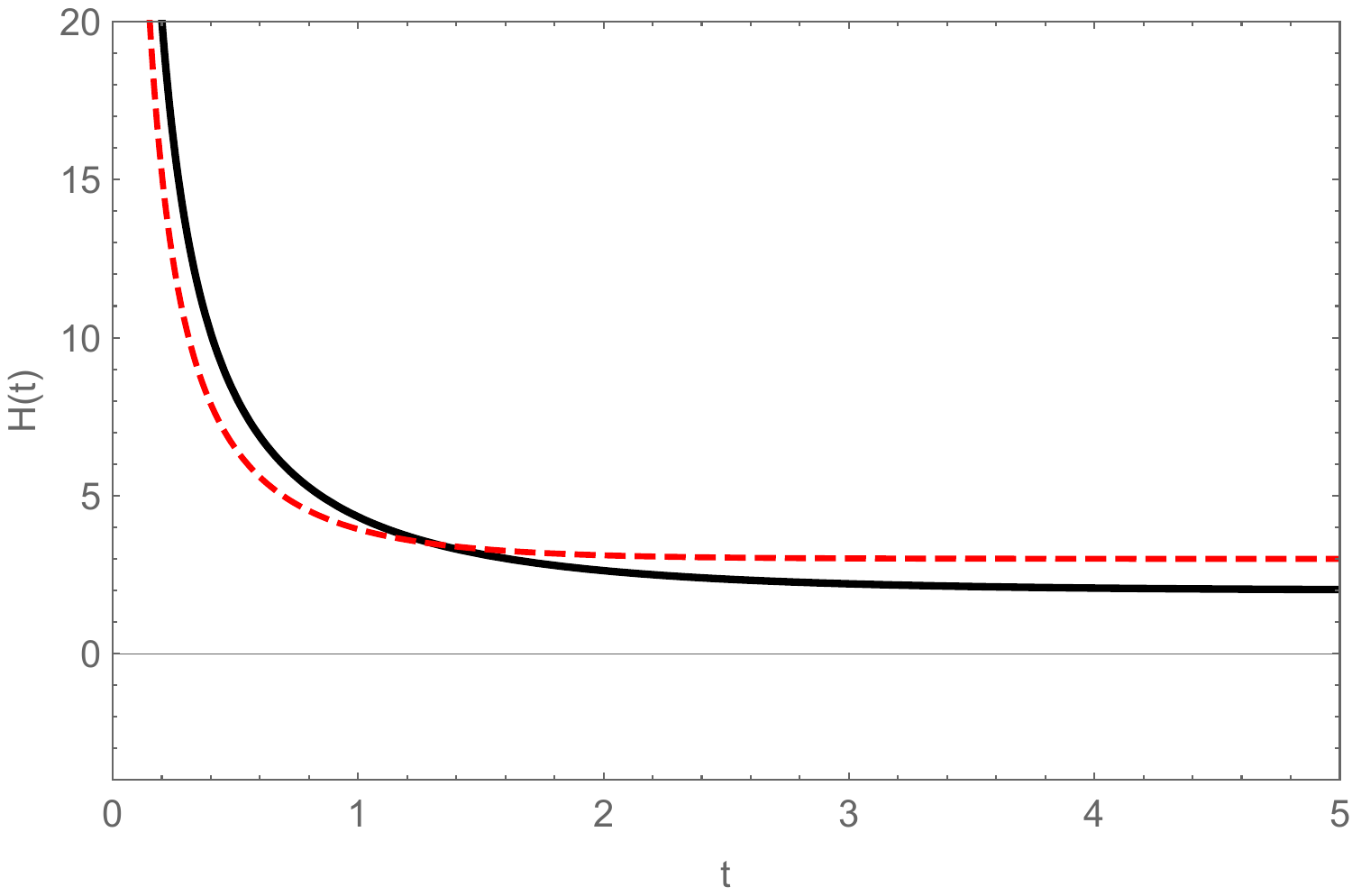}
\caption{Plots of the scale factor $a(t)$ (left panel) and Hubble's parameter $H(t)$ (right panel) for $A=2$, $B^2=1/4$ (thick line) and $A=3$, $B^2=1/3$ (dotted line).}\label{fig4}
\end{figure}

For this new case, we show in Fig.~\ref{fig4} the scale factor and Hubble's parameter for some choice of parameters. The scale factor for $B^2=1/n$ with $n$ even is always positive and increasing, and for $n$ odd, it is negative, decreasing, and in both cases the evolution is accelerated, with the Hubble parameter positive, decreasing asymptotically to a constant value. This model appears to evolve from a initial singularity, somehow similar to the Big Bang model. In late times, we can see from the Hubble parameter that the cosmic evolution leads to de Sitter geometry.

\begin{figure}[ht]
\centering
\includegraphics[{height=6cm,width=7cm,angle=00}]{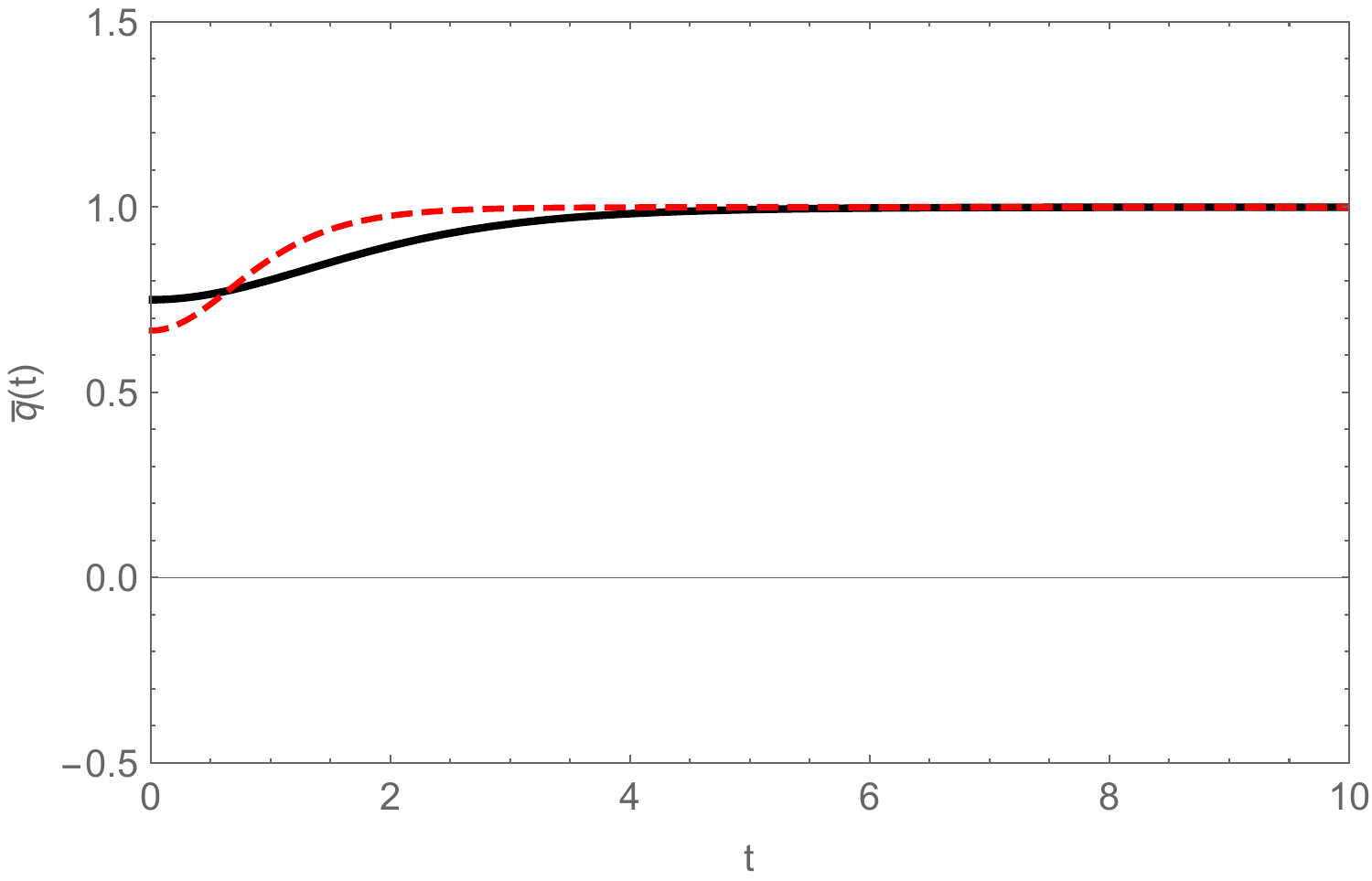}
\includegraphics[{height=6cm,width=7cm,angle=00}]{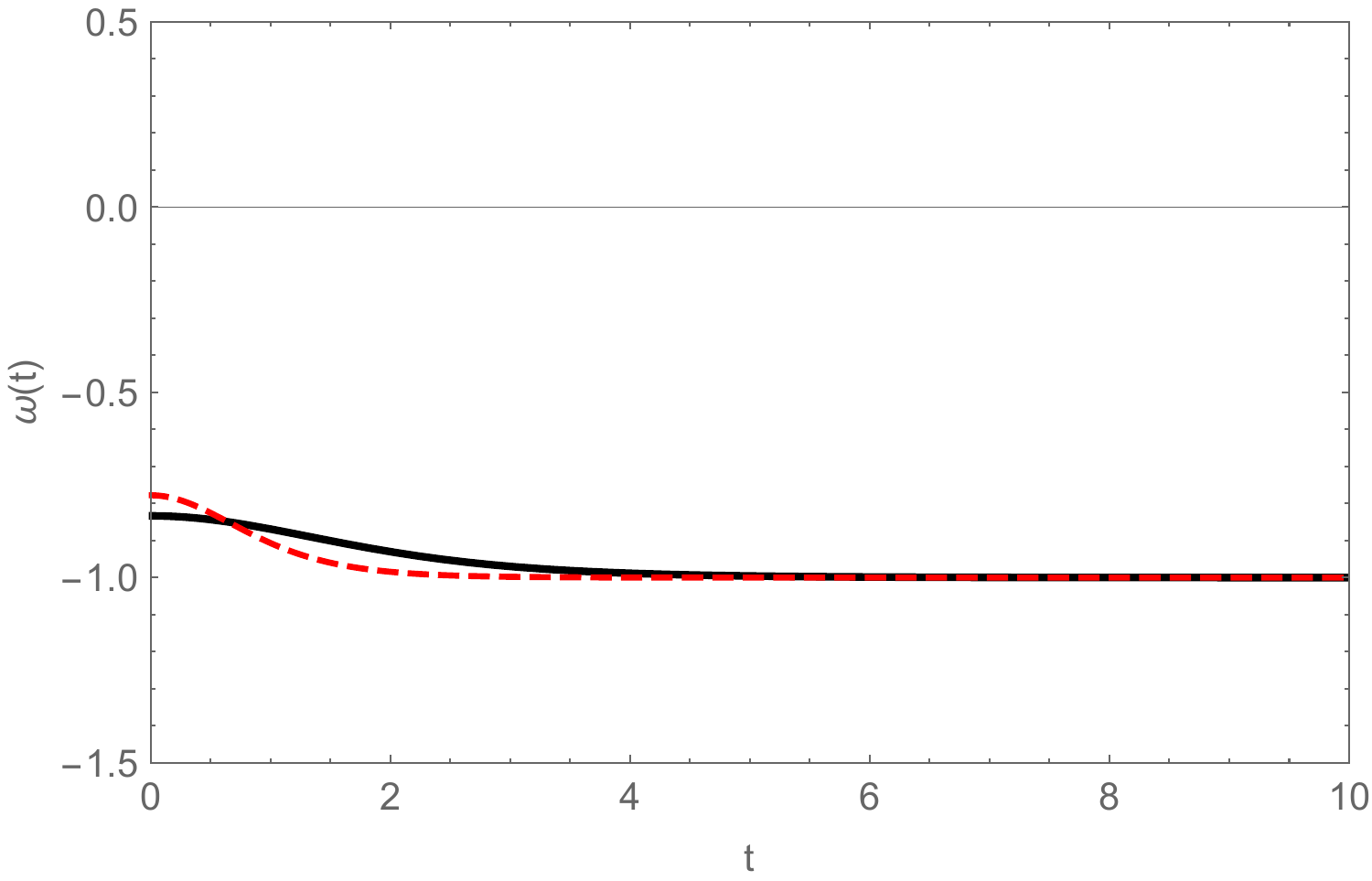}
\caption{The acceleration parameter $\bar{q}(t)$ (left panel) and $\omega$ (right panel) for $A=2$, $B^2=1/4$ (thick line) and for $A=3$, $B^2=1/3$ (dashed line).}\label{fig5}
\end{figure}

In this new case the acceleration parameter has the form
\begin{eqnarray}
\bar{q}=1-B^2\tanh{\left(\ln{\left(\tanh{\left(\frac{AB^2t}{2}\right)}\right)}\right)}^2
\end{eqnarray}
and the energy density e pressure are given by
\be
\rho(t)=\frac{3}{2}A^2\cosh{\left(\ln{\left(\tanh{\left(\frac{AB^2t}{2}\right)}\right)}\right)}^2\\
\ee
\ben
p(t)&=&A^2B^2\sinh{\left(\ln{\left(\tanh{\left(\frac{AB^2t}{2}\right)}\right)}\right)}^2-\nonumber\\
\nonumber\\
& &-\frac{3}{2}A^2\cosh{\left(\ln{\left(\tanh{\left(\frac{AB^2t}{2}\right)}\right)}\right)}^2
\een
In Fig.~\ref{fig5} we plot the acceleration and $\omega.$ We notice that $\omega$ varies in the interval $0<\omega<-1,$ going to $-1$ asymptotically,
as it happens to be the case of FLRW cosmology with cosmological constant \cite{cst}.

\subsection{Third example}


At last, we show a third example described by
\be
\label{W3}
W=A\ \textrm{sech} \left(B \phi-C \right)
\ee
where $A$, $B$ and $C$ are real parameters.  This choice of Eq. \eqref{W3} leads to the following potential
\begin{eqnarray}
V(\phi)&=&-\frac{1}{2} A^2 \left(B^2 \tanh(C-B \phi)-3\right)^2
\textrm{sech}(C-B \phi)^2
\end{eqnarray}
In this case the scalar field is not exactly invertible. However we perform a inversion of the function obtaining $\phi(t)$ numerically. The advantage of the first order formalism is to integrate numerically only the inversion of the scalar field once all the others parameters can be written in terms of scalar field. In other methods we have to integrate numerically a system of differential equations, a much more difficult task.  

For the present potential $V$ and $W$, the set of Eqs.~(\ref{em1}-\ref{em3}) are 
solved giving the following relation
\begin{equation}
\label{tphi3}
t=\frac{\cosh (C-B \phi)+\ln \left(\tanh \left(\frac{1}{2} (C-B \phi)\right)\right)}{A B^2}.
\end{equation}
The Hubble's parameter $H$ is given by
\begin{equation}
H(\phi)= A\ \textrm{sech}(B\phi-C).
\end{equation}
The scale factor becomes
\be
a(\phi)=\sinh (C-B\phi)^{1/B^2}.
\ee
In this new case the acceleration parameter has the form
\begin{equation}
\bar{q}=1-B^2 \tanh (C-B \phi)^2,
\end{equation}
and the energy density and pressure are given by
\be
\rho(\phi)=\frac{3}{2} A^2 \textrm{sech}(C-B \phi)^2,
\ee
\be
p(\phi)=\frac{1}{2} A^2 \left(2 B^2 \tanh(C-B \phi)-3\right)^2 \text{sech}(C-B \phi)^2.
\ee
These parameters are obtained in terms of time $t$ using the {\it NSolve} command of {\it Mathematica} to solve Eq. (\ref{tphi3}) and 
 write $\phi(t)$ numerically.
\begin{figure}[h]
\centering
\includegraphics[{height=6cm,width=7cm,angle=00}]{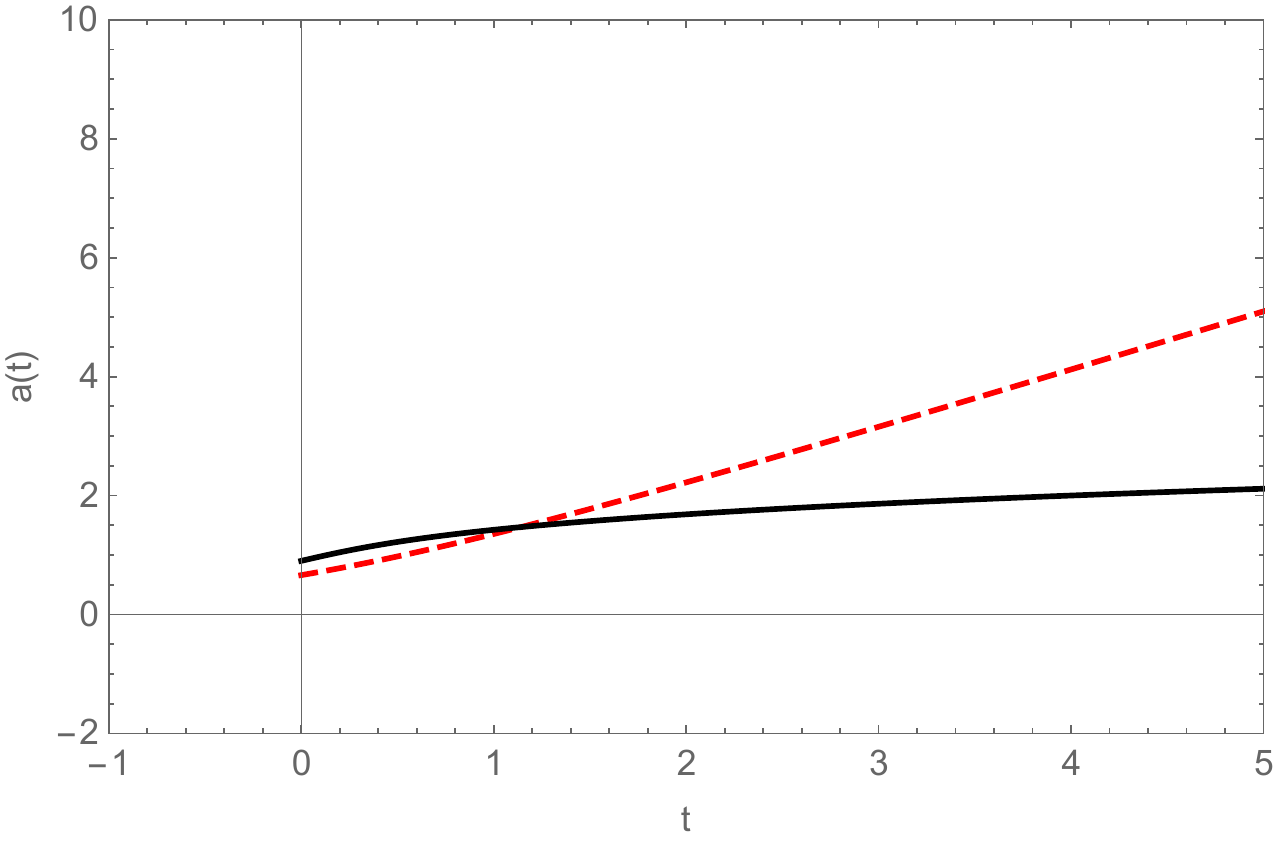}
\includegraphics[{height=6cm,width=7cm,angle=00}]{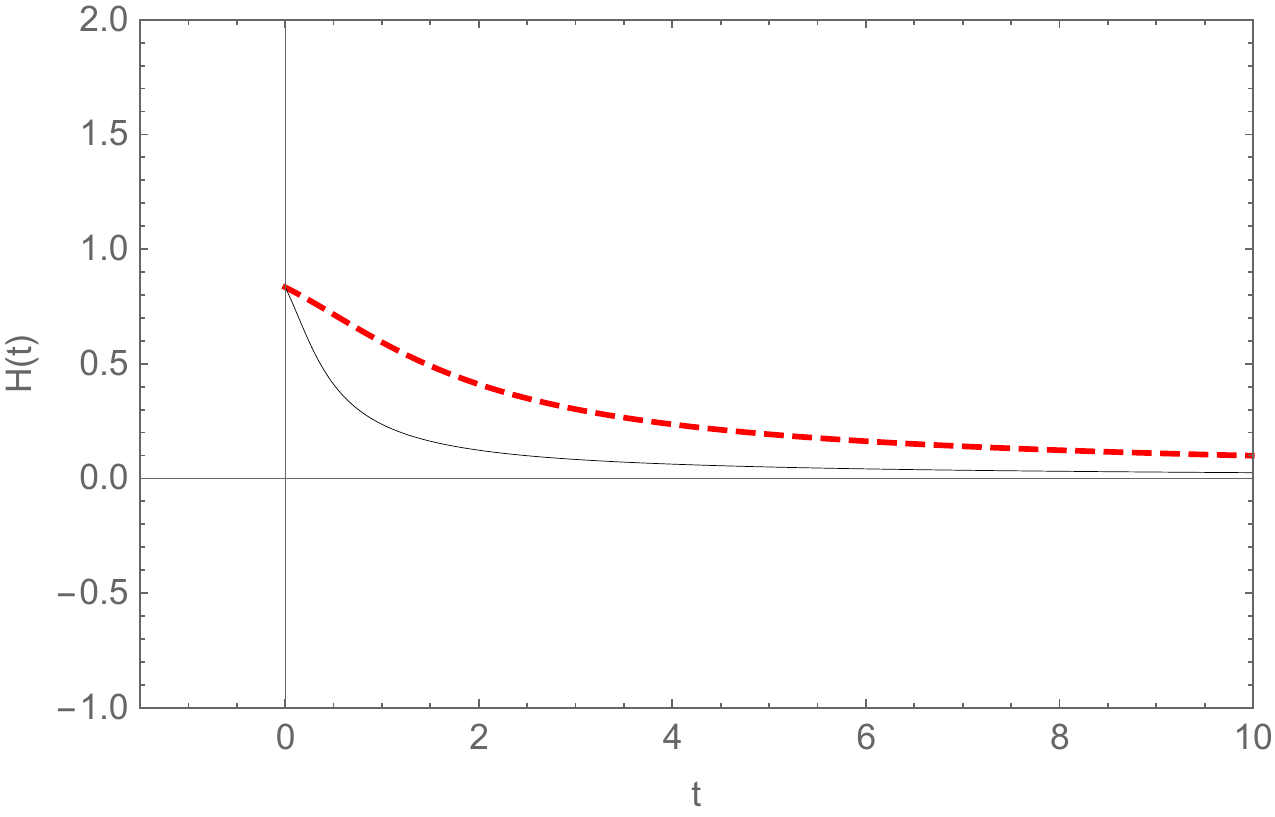}
\caption{Plots of the scale factor $a(t)$ (left panel) and Hubble's parameter $H(t)$ (right panel) for $A=1$, $B=-2$, $C=2$ (thick line) and $A=1$, $B=1$, $C=1$ (dashed line).}\label{fig6}
\end{figure}
These results are shown in the Figs. \ref{fig6} and \ref{fig7}.
Once more, we notice by the scale factor and Hubble's parameter that this model leads to expanding universes with decreasing rates of evolution depending on the values of parameters. In both cases the universes starts their evolution with minimum size, i.e., non-Big Bang model of universe. 
\begin{figure}[h]
\centering
\includegraphics[{height=6cm,width=7cm,angle=00}]{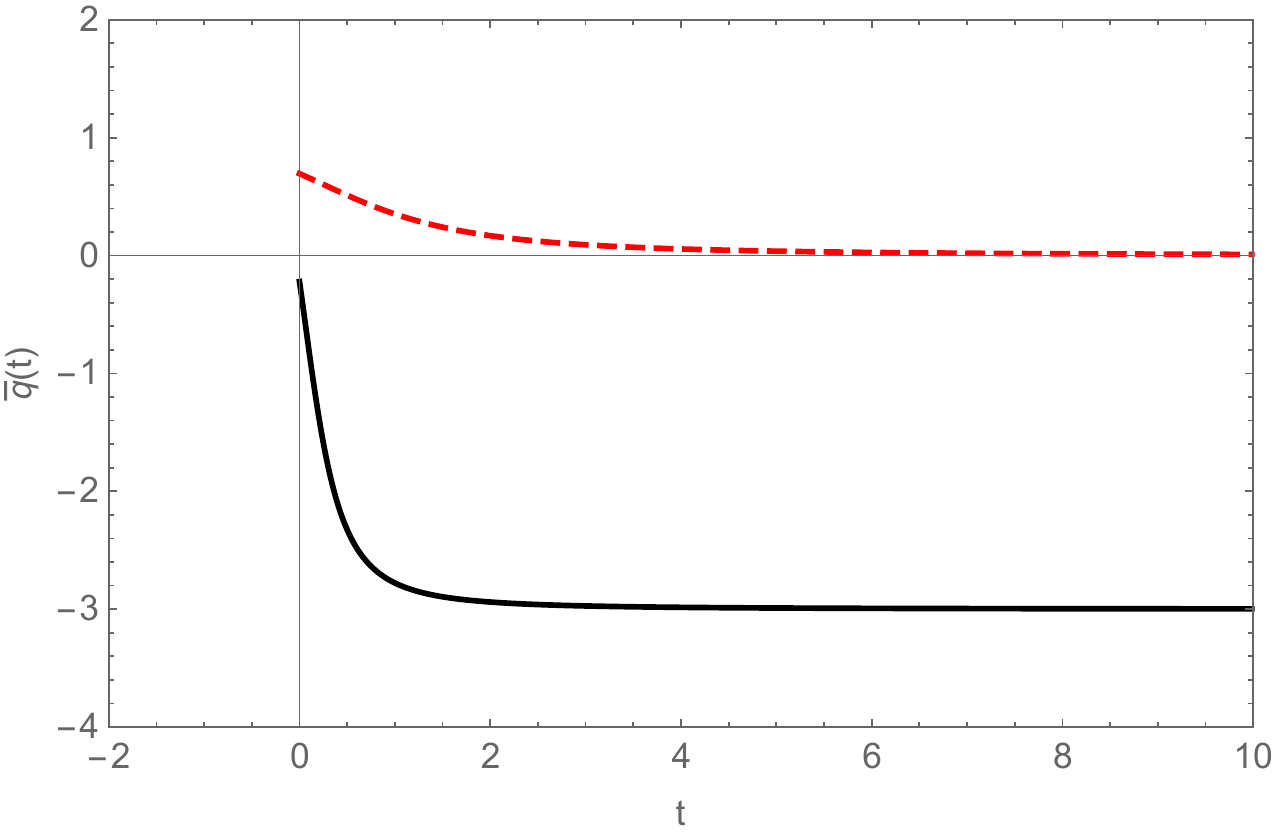}
\includegraphics[{height=6cm,width=7cm,angle=00}]{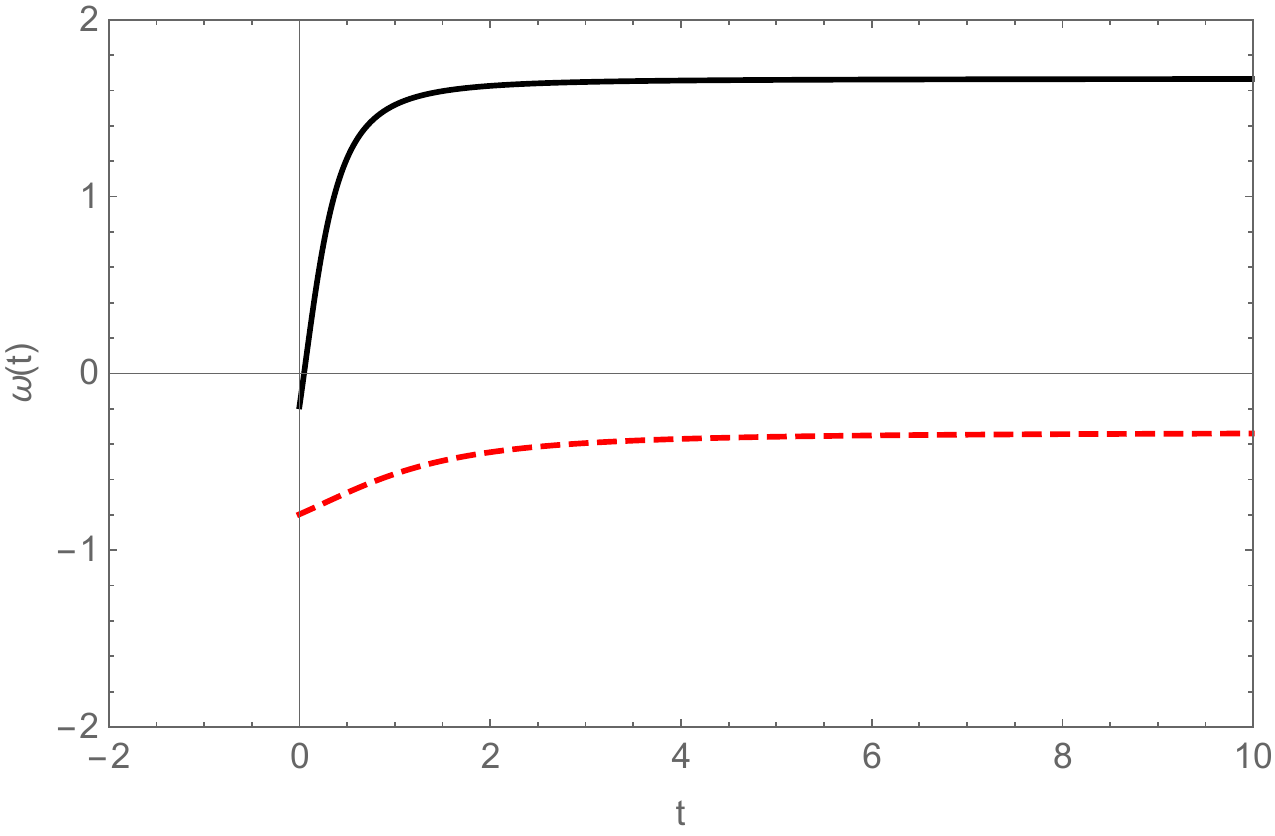}
\caption{The acceleration parameter $\bar{q}(t)$ (left panel) and $\omega$ (right panel) for for $A=1$, $B=-2$, $C=2$ (thick line) and $A=1$, $B=1$, $C=1$ (dashed line).}\label{fig7}
\end{figure}
The $\omega$ can assume positive and negative values approximating to a constant value when $t$ increases.

\section{The deformation procedure}

Now we turn our attention to the deformation procedure introduced in Ref.~\cite{blm}. The idea is to
bring the methodology behind the deformation procedure to FLRW cosmology, with the motivation that since the deformation change the original scalar field model, in the cosmological scenario it will also change the cosmic evolution. The approach may then be used to find families of soluble cosmological models, driven by a single real scalar field.

We follow the original work \cite{blm}, but now we start with the Einstein-Hilbert action (\ref{model}). We deform the scalar field using $\phi
=f({\wt\phi})$, with $f$ being the deformation function, a real function of the new field $\wt\phi.$ We suppose that $f:{\bf R}\to {\bf R}$ is bijective, thus invertible and 
differentiable.

For each model represented by the potential $V(\phi)$, giving by (\ref{pot}) and having solutions $\phi(t)$ and $H(t)$ satisfying Eqs.~(\ref{be0}) and (\ref{bek}), and thus 
Eqs.~(\ref{em1}-\ref{em3}), we can introduce another model, the deformed model, which is described by the deformed scalar field and is defined by
\be
{\wt V}({\wt \phi})=\frac32{\wt W}^2-\frac12{\wt W}_{{\wt \phi}}^2
\ee
with
\be 
{\wt W}=\int_0^{{\wt \phi}}\frac{W_\phi[\;f({\wt \phi})\;]}{df/d{\wt \phi}}{d{\wt \phi}}
\ee

The key point of the deformation procedure is that if $\phi(t)$ and $H(t)=W(\phi(t))$ solve the original model, then the deformed model is solved by 
${\wt\phi}(t)=f^{-1}[\;\phi(t)\;]$ and ${\wt H}(t)={\wt W}[{\wt\phi}(t)]$, for $f^{-1}$ being the inverse of $f.$

We illustrate the above result with deformations of the examples discussed in the section \ref{s2}.

\subsection{First deformed example}

Firstly, we consider the model defined by Eq.(\ref{W1}) and the deformation function $f(\tilde{\phi})=\ln{\tilde{\phi}}$. This 
leads to a new model described by the deformed potential
\begin{eqnarray}
\widetilde{V}&=&\frac{3n^2}{2}\left[\widetilde{W}_{1}\ln{(\tilde{\phi})}^{n-1}-\left(\frac{n-1}{2}\right)\widetilde{W}_{n-1}\right]^2+
\nonumber\\
& &\;-\frac12{A^2n^2\ln{(\tilde{\phi})}^{2n-2}}
\end{eqnarray}
where
\begin{eqnarray}
\widetilde{W}_{n}=n\left[\widetilde{W}_{1}\ln{(\tilde{\phi})}^{n-1}-\frac{n-1}{2}\widetilde{W}_{n-1}\right]+C_{n}
\end{eqnarray}
for $n\geq2,$ and for $n=1$
\begin{equation}
\widetilde{W}_{1}=\frac12{A\tilde{\phi}^2}
\end{equation}
For this model, from Eq.(\ref{edc}) the deformed field is obtained as $\wt\phi=\exp(\phi)$. Thus, we can use Eq.~(\ref{edc}) to write
\begin{eqnarray}
\tilde{\phi}_{n}(t)&=&e^{[An(n-2)t]^{\frac{1}{2-n}}}\qquad n\neq2\\
\nonumber\\
\tilde{\phi}_{2}(t)&=&e^{e^{-2At}}\qquad n=2
\end{eqnarray}
The Hubble parameter has the following form
\begin{eqnarray}
\widetilde{H}_{n}(t)&=&\frac12 {n}{A e^{2[An(n-2)t]^{\frac{1}{2-n}}}}\left[An(n-2)t\right]^{\frac{n-1}{2-n}}+ \nonumber\\
& &-\frac12 n(n-1)\widetilde{W}_{n-1}+C_{n}\;\;\;\; n\neq2
\end{eqnarray}
and
\be
\widetilde{H}_{2}(t)=A e^{2e^{-2At}}\left(e^{-2At}-\frac12\right)+C_{2}\;\;\; n=2
\ee
and the acceleration parameter gets to
\be
{\wt{\bar{q}}}_{2}(t)=1-\left(\frac{2A\tilde{\phi}
\ln(\tilde{\phi})}{{A\tilde{\phi}^2}(\ln(\tilde{\phi})-1)+C_{2}}\right)^2\;\;\;\;for\ \ n=2
\ee
and
\be
{\wt{\bar{q}}}_{n}(t)=1-n^2\frac{N_1^2}{N_2^2}\;\;\;\;for\ \ n\neq 2
\ee
where we have set
\ben
N_1&=&2\tilde{\phi}A\ln(\tilde{\phi})^{(n-1)}+(n-1){A\tilde{\phi}}\ln(\tilde{\phi})^{(n-2)}+
\nonumber\\
& & - (n-1) \frac{ d\widetilde{W}_{n-1} }{ d\tilde{\phi} }
\een
and
\be
N_2={n{A\tilde{\phi}^2}\ln(\tilde{\phi})^{(n-1)}-{n(n-1)\widetilde{W}_{n-1}}+2\;C_{n}}
\ee
\begin{figure}[ht]
\centering
\includegraphics[{height=6cm,width=7cm,angle=00}]{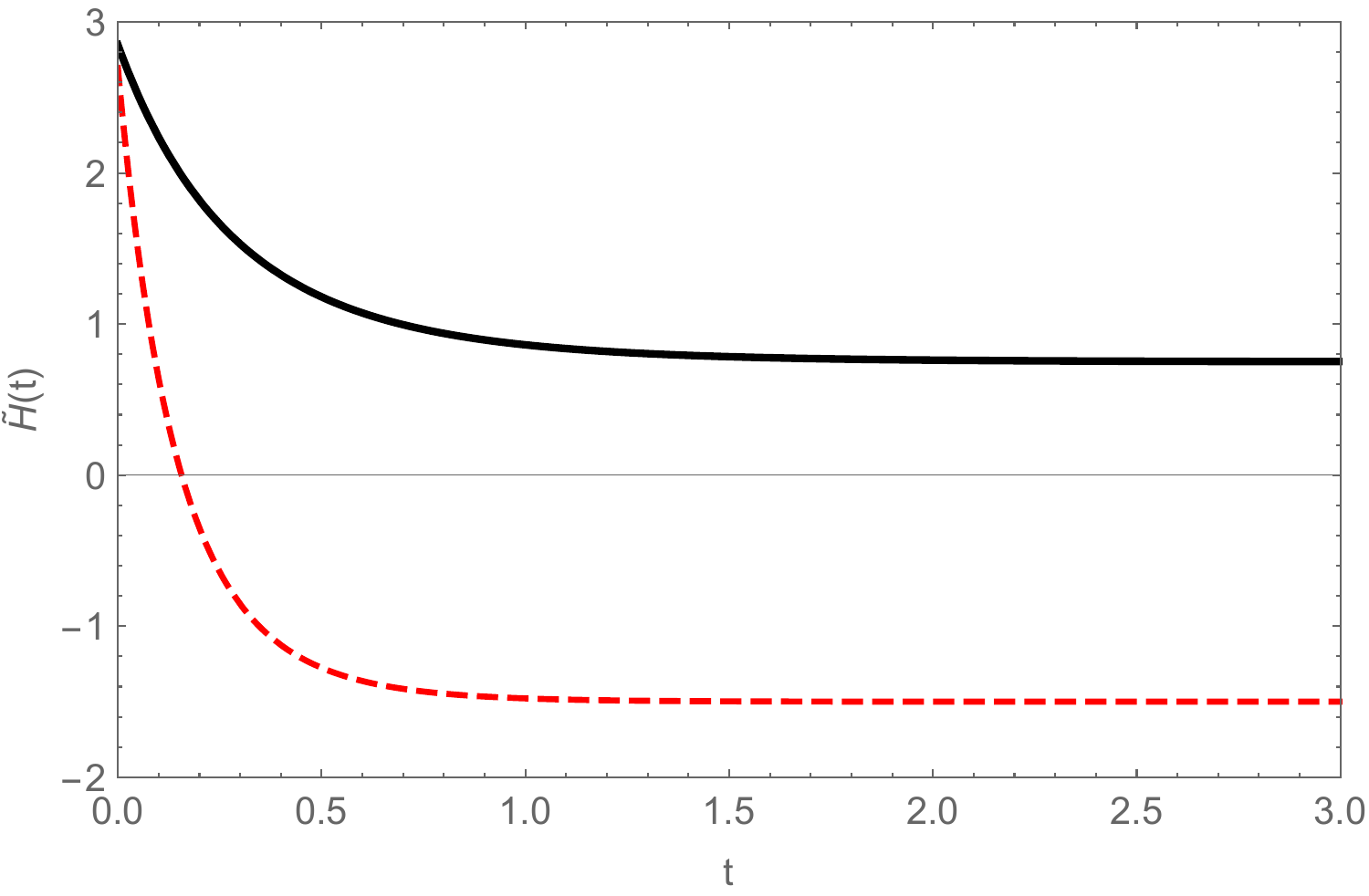}
\includegraphics[{height=6cm,width=7cm,angle=00}]{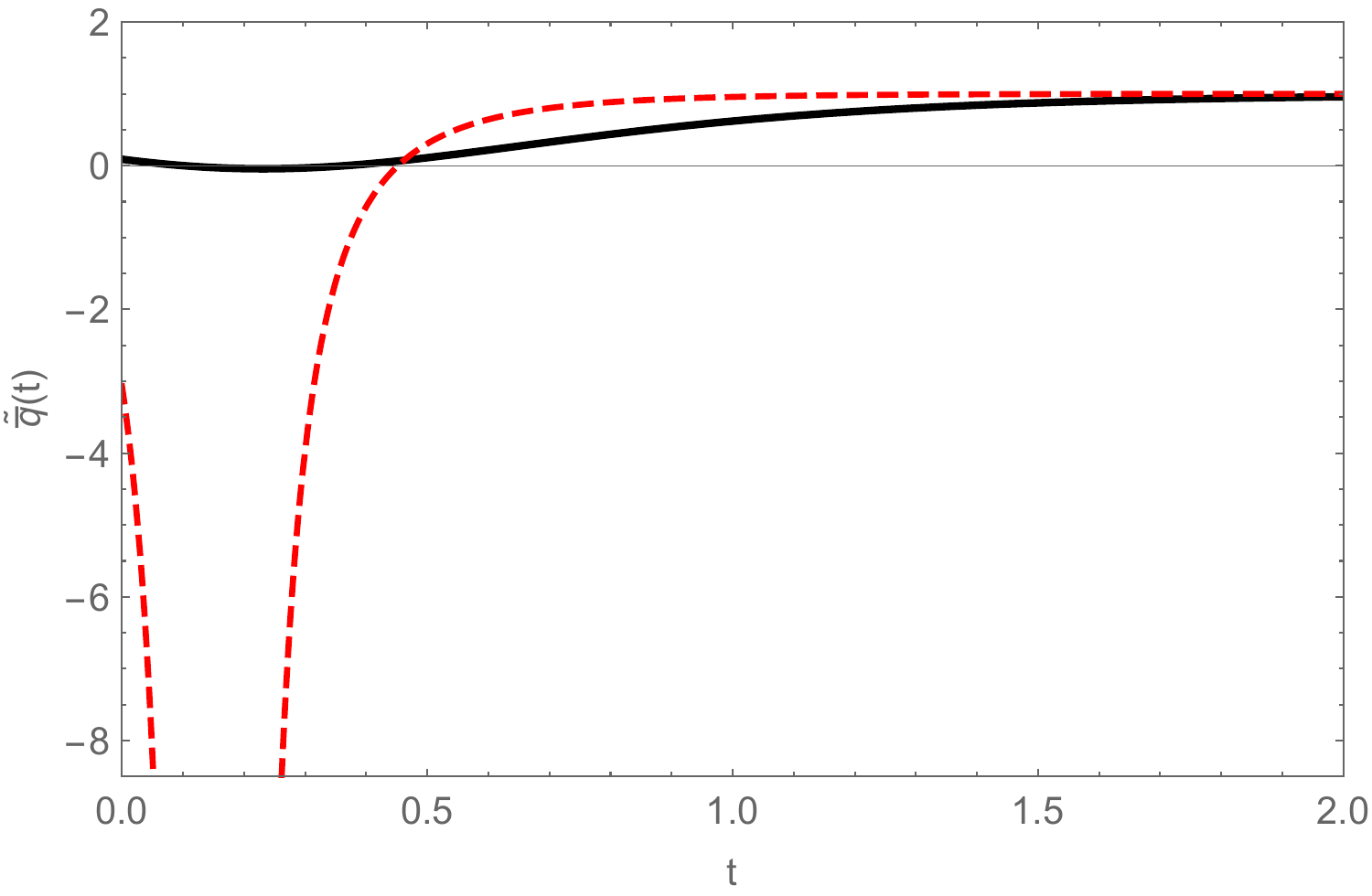}
\caption{Plots of the Hubble's parameter ${\wt H}(t)$ (left panel) and the
acceleration parameter ${\wt{\bar{q}}}(t)$ (right panel) for $A=1$, $C_{2}=-1$ (dashed line), and for $A=1/2$, $C_{2}=1$ (thick line).}
\label{fig8}
\end{figure}

In this case, we could not obtain analytic expressions for the scale factor and so we use ${\wt H}$ and ${\wt{\bar{q}}}$ to analyze the above cosmic
evolution. They are plotted in Fig.~\ref{fig8} and the constant $C_{2}$ can be used to make ${\wt{\bar{q}}}(t)$ divergent. 
\subsection{Second deformed example}

Here we introduce another example, described by the function $W=A\cosh(B\phi)+D$. And also, we use the deformation function 
$f(\tilde{\phi})=\frac{1}{B}\ln{(\tanh{(E\sinh{(B\tilde{\phi})})})}.$ $A,$ $B,$ $D,$ and $E$ are real parameters. In this case the new model is represented by the deformed 
potential
\begin{eqnarray}
\widetilde{V}\!=\!\frac{3}{2}\left(D-\frac{A}{E}{\rm
arctan}(e^{B\tilde{\phi}})\right)^2\!\!-\frac{1}{8}\frac{A^2B^2}{E^2\
\cosh^2{(B\tilde{\phi})}}
\end{eqnarray}
and from the inverse of the deformation function we have
\begin{eqnarray}
\tilde{\phi}=\frac{1}{B} {\rm arcsinh}\left(\frac{AB^2t}{2E}\right)
\end{eqnarray}
\begin{figure}[ht]
\centering
\includegraphics[{height=6cm,width=8cm,angle=00}]{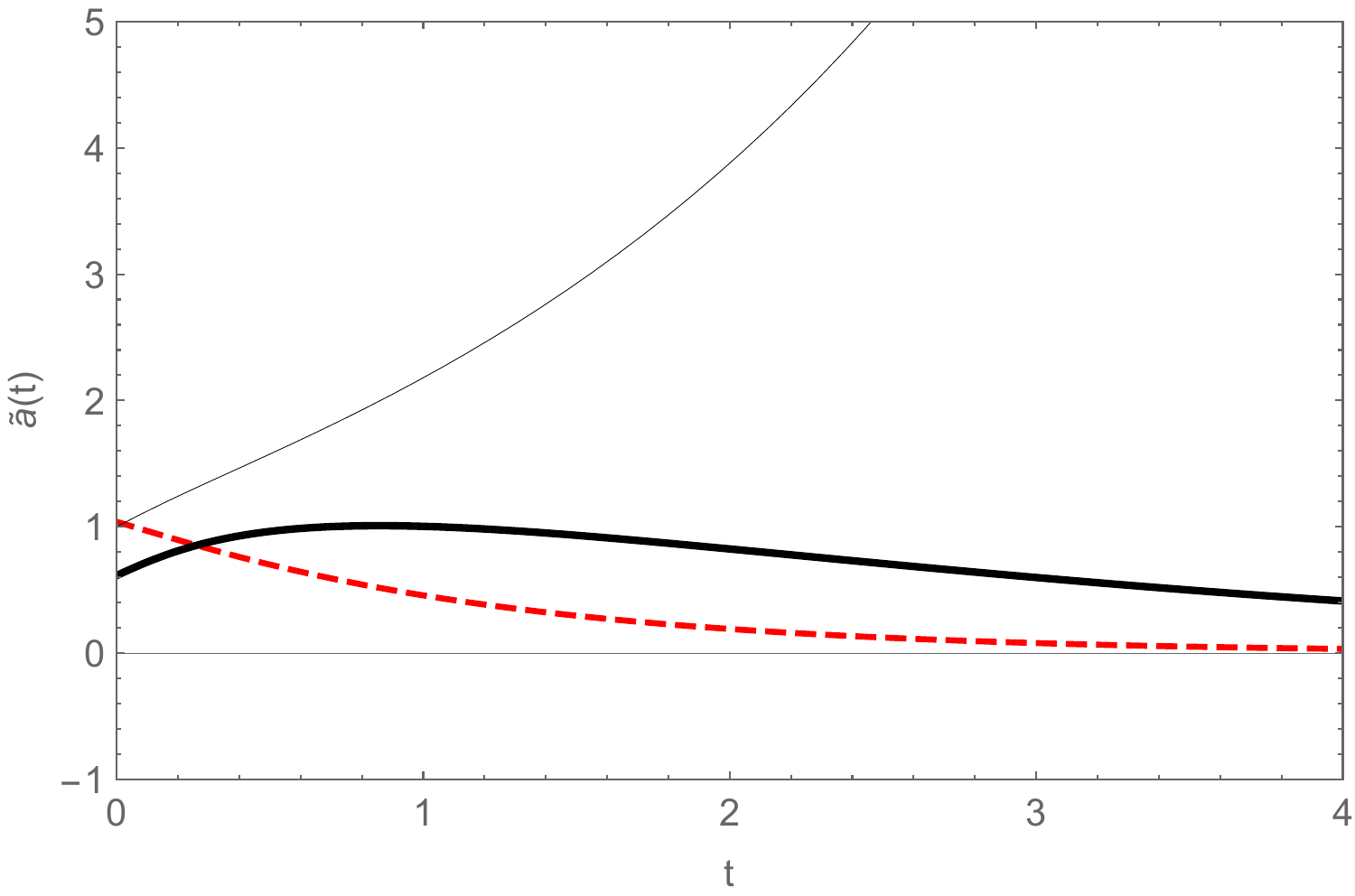}
\includegraphics[{height=6cm,width=8cm,angle=00}]{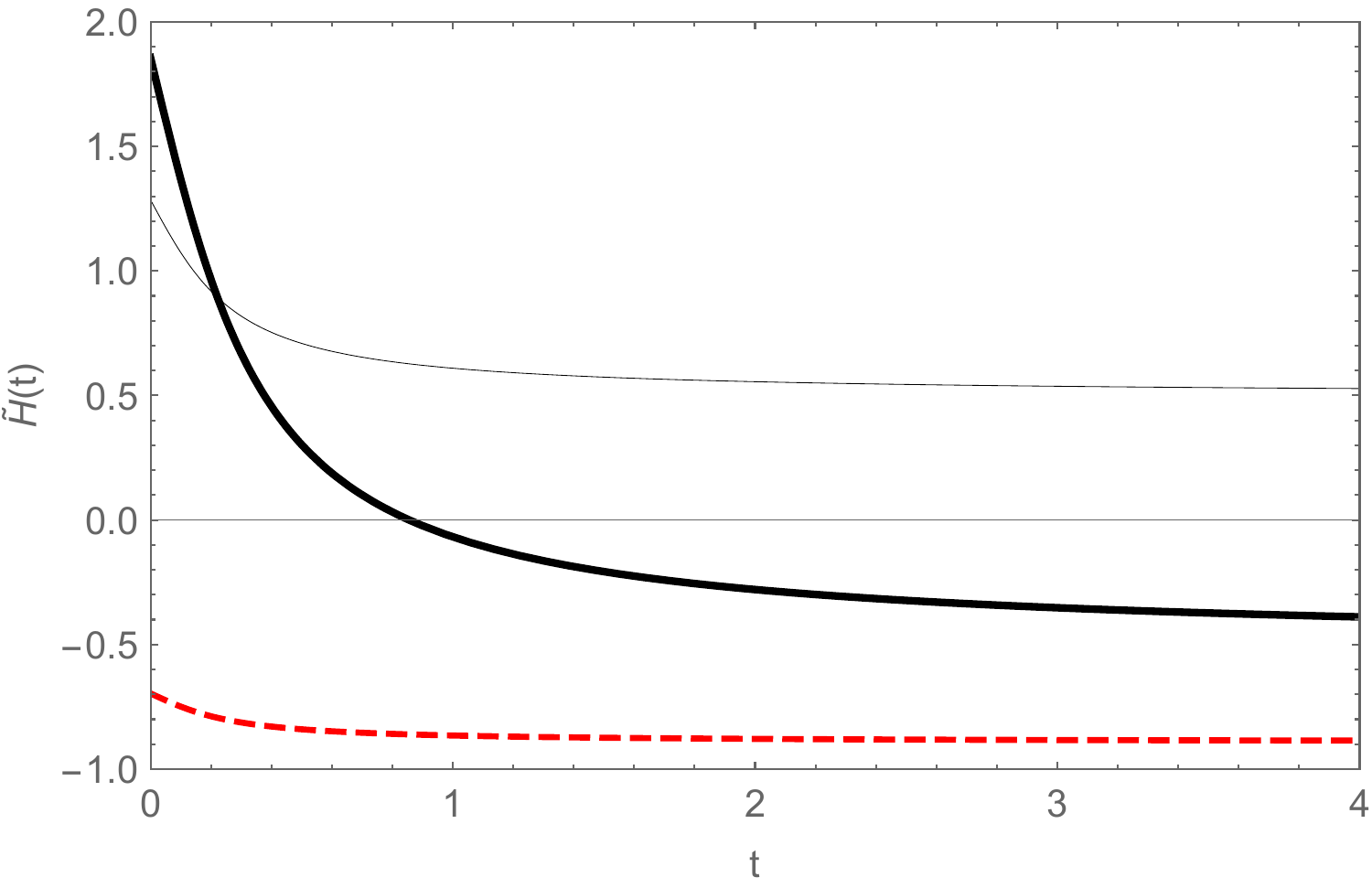}
\includegraphics[{height=6cm,width=8cm,angle=00}]{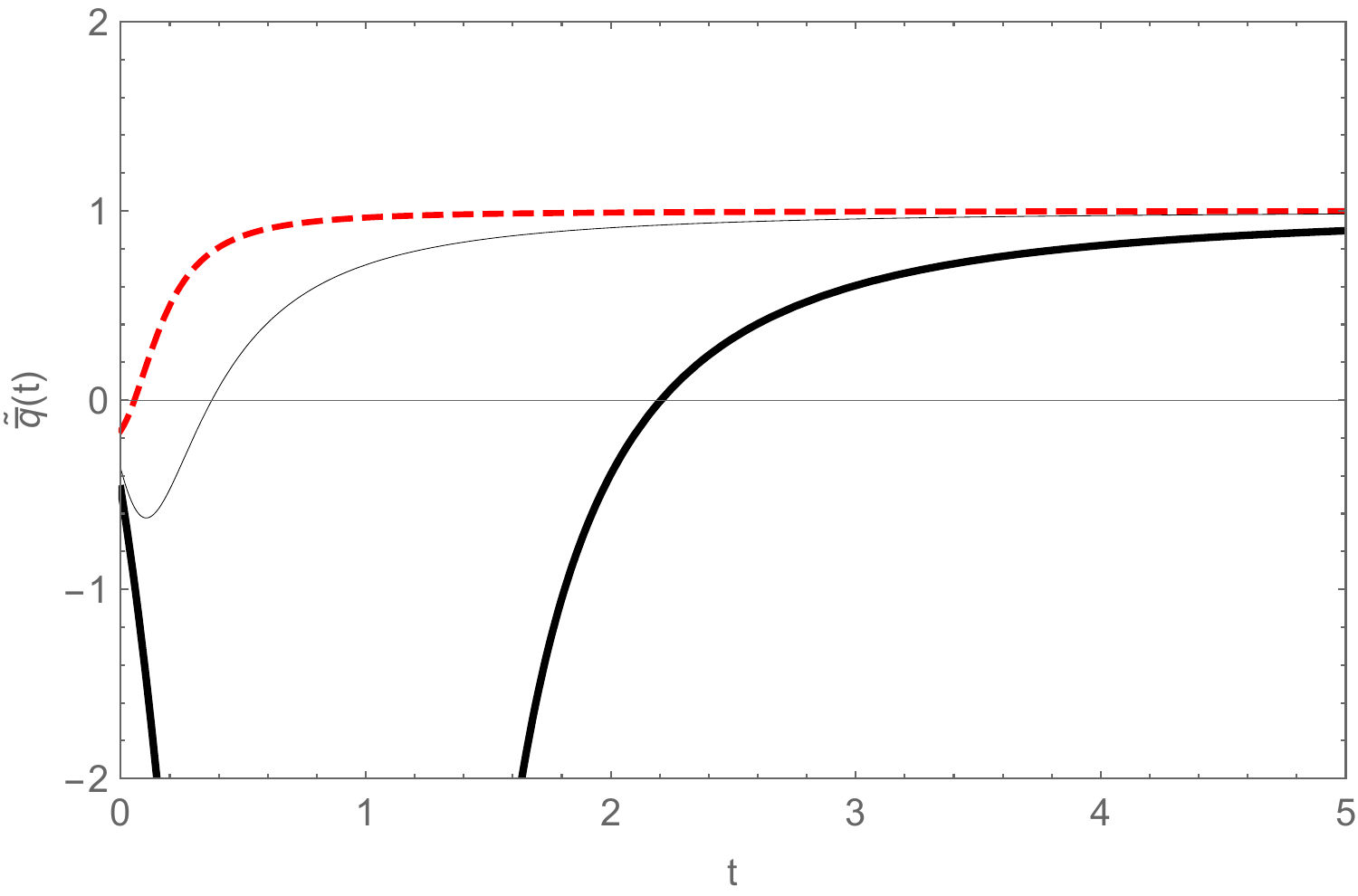}
\caption{The scale factor ${\wt a}(t)$ (top panel), the Hubble's parameter ${\wt H}(t)$ (center panel) and acceleration parameter ${\wt{\bar{q}}}(t)$ (bottom panel), for $A=-1/2$, $B=3$, $D=1/2$, $E=1/2$ (thin line), $A=-1/2$, $B=3/2$, 
$D=-1/2$, $E=1/6$ (thick line), $A=1/2$, $B=6$, $D=-1/2$, $E=2$ (dashed line).}
\label{fig9}
\end{figure}

The deformed Hubble parameter is given by
\be
\tilde{H}(t)=-\frac{A}{E}{\rm \arctan}\left(\frac{AB^2t+E\sqrt{4E^2+A^2B^4t^2}}{2E}\right)+D
\ee
and the deformed scale factor is
\be
\tilde{a}(t)=N_3^{\frac1{B^2}}\exp\left(-\frac{At}{E}{\rm \arctan}\left(\frac{AB^2t+N_3}{2E}\right)\!\!+\!\!Dt\right)
\ee
where we have set
\be
N_3=(4E^2+A^2B^4t^2)^{1/2}
\ee
In the present case, the acceleration parameter gets to the form
\ben
{\wt{\bar{q}}}=1-\frac{A^2B^2}{4E^2}\,\biggl[ N_4\,\cosh{\left({\rm arcsinh}\left(\frac{AB^2t}{2E}\right)\right)}\biggr]^{-2}
\een
where we have set
\be
N_4=-\frac{A}{E}{\rm \arctan}\left(e^{{\rm
arcsinh}\left(\frac{AB^2t}{2E}\right)}\right)+D
\ee
Also, the energy density and pressure are given by
\begin{eqnarray}
\tilde{\rho}(t)&=&\frac{3}{2}\left(D-\frac{A}{E}{\rm arctan}\left(e^{{\rm arcsinh}\left(\frac{AB^2t}{2E}\right)}\right)\right)^2\\
\nonumber\\
\tilde{p}(t)&=&\frac14\frac{A^2B^2}{E^2\cosh^2{\left({\rm arcsinh}\left(\frac{AB^2t}{2E}\right)\right)}}-\tilde{\rho}(t)
\end{eqnarray}
In Fig.~\ref{fig9} we plot the scale factor ${\wt a}(t)$, the Hubble parameter ${\wt H}(t)$ and the acceleration parameter ${\wt{\bar{q}}}(t)$, for some choice of parameters, with $D\neq0.$ We see that the behavior of the scale factor follows three generic possibilities: increasing, increasing-decreasing, and decreasing.
The Hubble parameter, however, decreasing in these cases, getting to a positive or negative constant asymptotically. This model has an
interesting feature, related to the many distinct scenarios available from the many specific choice of parameters. We notice,
however, that there is no cosmic evolution with initial singularity, as it happens with the Big Bang model.

\subsection{Third deformed example}
As a third example of the deformation procedure we convert an expanding universe in a static one. We start recovering the function $W=A\ \textrm{sech}(B\phi-C)$ and using the deformation function,
\be
f(\tilde{\phi})=\frac{C}{B}+\frac{\textrm{arctanh}(D \tilde{\phi})}{B}
\ee
being $A,$ $B,$ $C,$ and $D$ real parameters. The only new parameter introduced by the deformation is $D$. In this case the new model is represented by the new potential given by
\begin{equation}
\widetilde{V}=\frac{1}{50} A^2 B^4 \left(1-D^2 \tilde{\phi} ^2\right)^3 \left(\frac{3 \left(D^2 \tilde{\phi}^2-1\right)^2}{D^4}-25\ \tilde{\phi} ^2\right)
\end{equation}
and from the inverse of the deformation function we have
\be
\tilde{\phi}= \frac{\tanh (B\phi-C )}{D}.
\ee
Once we do not have the function $\phi(t)$ exactly, it is important to set the parameters $A,$ $B,$ and $C$ as the same values used in the numerical calculation. The parameter $D$ can be chosen freely.

\begin{figure}[h]
\centering
\includegraphics[{height=6cm,width=8cm,angle=00}]{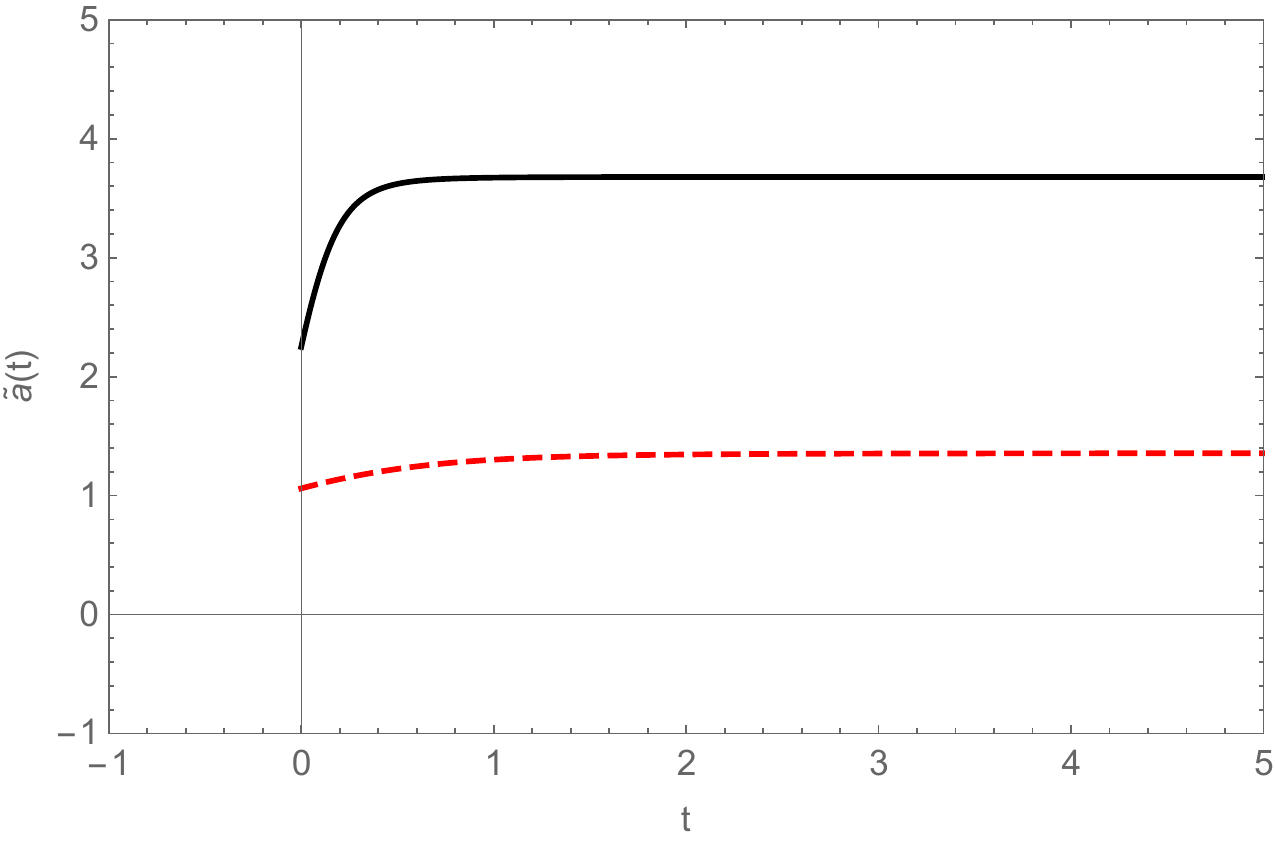}
\includegraphics[{height=6cm,width=8cm,angle=00}]{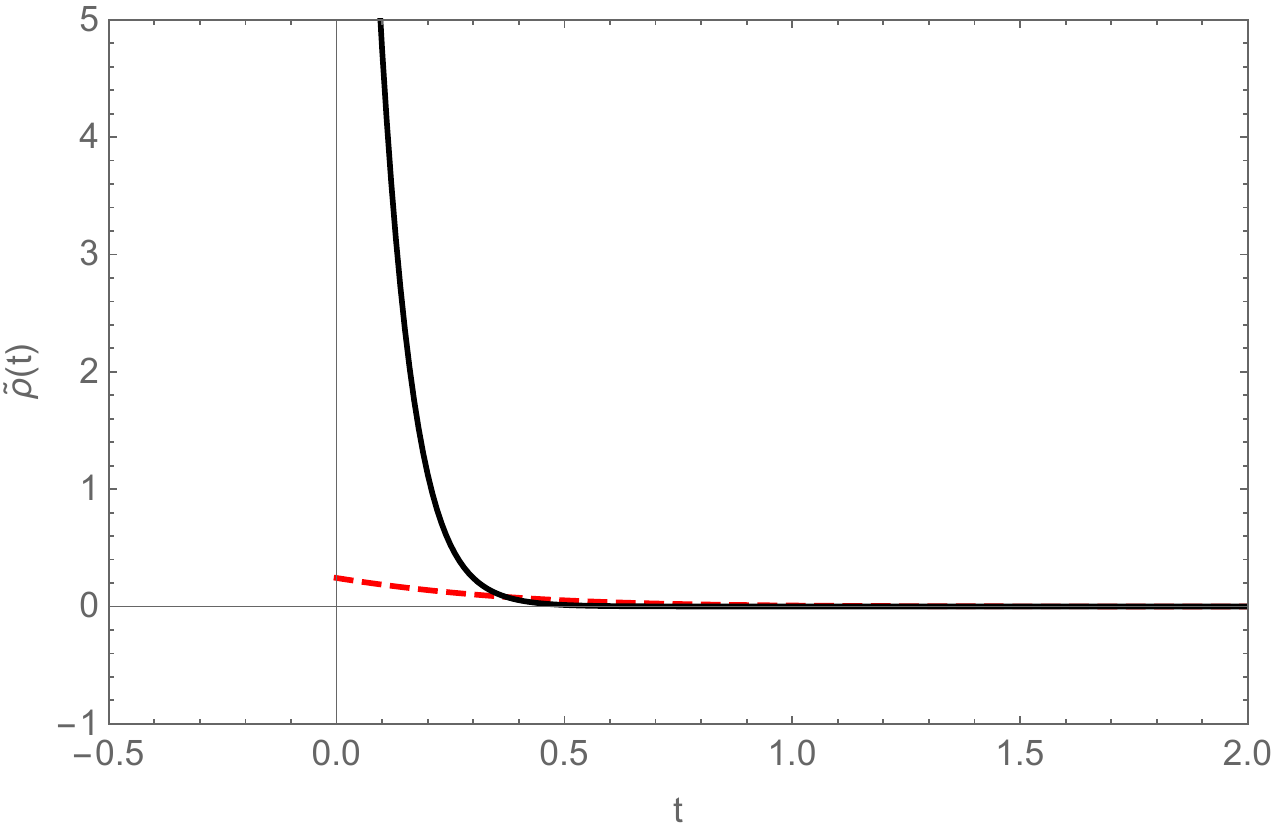}
\includegraphics[{height=6cm,width=8cm,angle=00}]{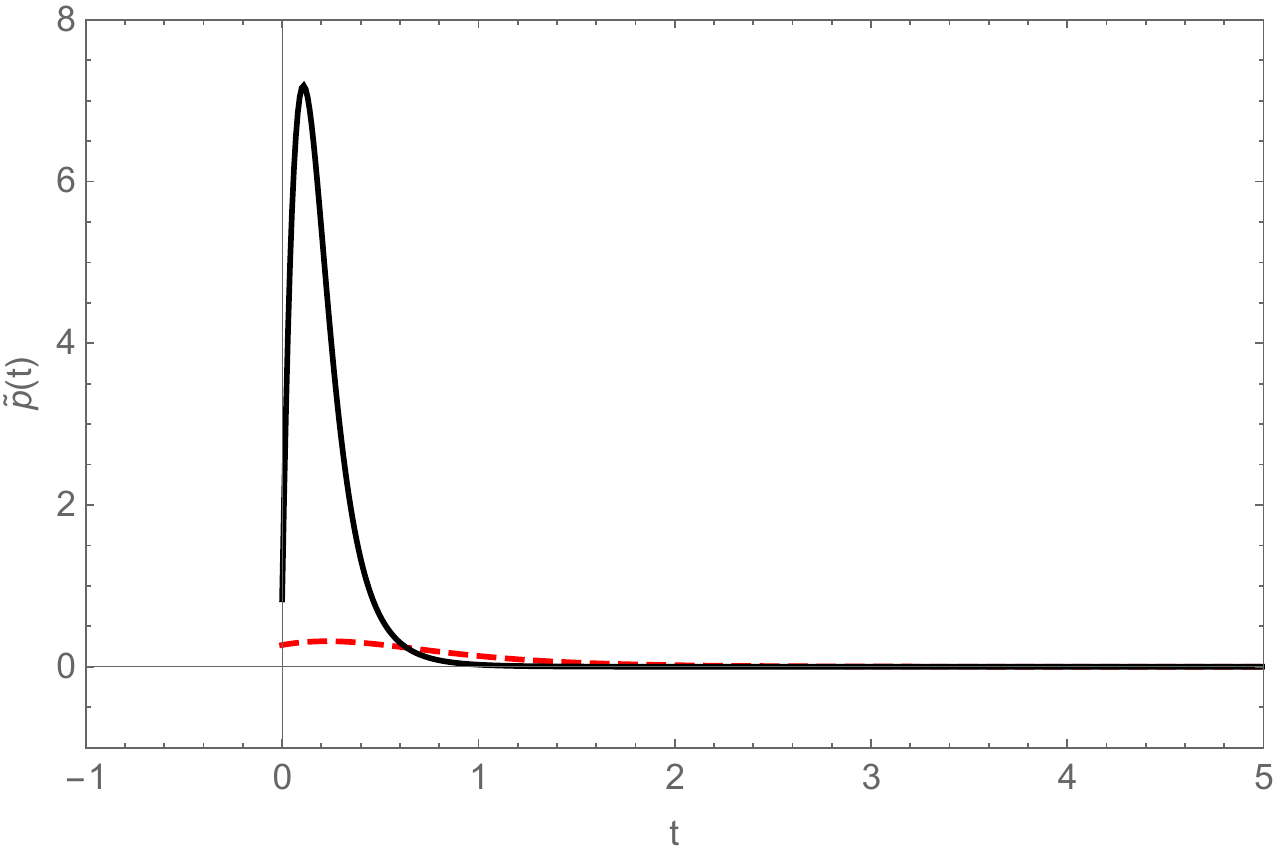}
\caption{The scale factor ${\wt a}(t)$ (top panel), the energy density ${\wt \rho}(t)$ (center panel) and pressure ${\wt{p}(t)}$ (bottom panel), for $A=1$, $B=-2$, $C=2$
$D=-1/\sqrt{10}$ (thick line), $A=1$, $B=1$, $C=1$ and $D=-1/\sqrt{5}$ (dashed line).}
\label{fig10}
\end{figure}

The deformed Hubble parameter is given by
\be
\tilde{H}(\tilde{\phi})=\frac{A B^2 \left(1-D^2 \tilde{\phi} ^2\right)^{5/2}}{5D^2}
\ee
and the deformed scale factor is
\be
\tilde{a}(\tilde{\phi})=\exp\left(-\frac{\tilde{\phi}^2}{10}\right)\ \tilde{\phi} ^{\frac{1}{5D^2}}
\ee
In the present case, the energy density and pressure are given by
\begin{eqnarray}
\tilde{\rho}(\tilde{\phi})&=&\frac{3 A^2 B^4 \left(1-D^2 \tilde{\phi} ^2\right)^5}{50D^4}\\
\nonumber\\
\tilde{p}(\tilde{\phi})&=&\frac{1}{50}A^2 B^4 \left(1-D^2 \tilde{\phi} ^2\right)^3 \left(50\ \tilde{\phi} ^2-\frac{3 \left(D^2 \tilde{\phi}^2-1\right)^2}{D^4}\right)
\end{eqnarray}
These quantities can be expressed in terms of the time $t$ using the numerical results obtained for $\phi(t)$. We plot, in Fig.~\ref{fig10}, the scale factor ${\wt a}(t)$, the energy density ${\wt \rho}(t)$ and the pressure ${\wt{p}}(t)$, for some choice of parameters, with $D<0.$ We see that the scale factor expands until a constant values, i.e., a static universe when $t$ increases. Additionally we notice that there is no initial singularity, as it happens with the Big Bang model. The energy density and pressure go fast to small values. An universe filled with diluted dust has such characteristics.

\section{Ending comments}

In this work we have revised and included new examples of the first-order formalism put forward in Ref.~\cite{bglm}. We did this in the simpler case of flat space-time, to provide the cosmological scenario needed to implement our key idea, of bringing to the cosmological scenario the methodology behind the deformation procedure put forward in
Ref.~\cite{blm} in the absence of gravity.

The main results of the present work show that the deformation procedure can be nicely implemented for FLRW models driven by scalar field under the first-order formalism, which 
helps significantly the search for analytical solutions. As we have shown, the deformation procedure introduced in this work may lead to new and interesting scenarios, obtained 
from simpler models of FLRW cosmology. In general, we can deform an expanding model into
another, expanding or contracting model. These possibilities depend not only on the deformation function, but also on the parameters which control the model to be deformed.

The present methodology may be used to investigate several interesting issues, in particular the case in which the cosmic evolution occurs in closed, flat or open geometry, for scalar field evolving under standard or tachyonic dynamics. Another issue concerns the use of the deformation procedure for models in which the equations of motion cannot be reduced to first-order equations. The investigations will certainly enlighten the relation between the deformation procedure and the approaches introduced in Refs.~{\cite{barrow95,sahni}}. 
Evidently, the construction of new models with the deformation procedure developed in this work may open a novel possibility to discover interesting scalar field models in which the attractor behavior is the field oscillating indefinitely with finite amplitude, as required to describe the new cosmological phase proposed in Ref. \cite{wil}, sometimes referred to as time crystals \cite{TC1,TC}.
Another possibility is to consider the addition of other scalar fields. This was developed before in flat spacetime in Ref. \cite{sala}, for instance, and we can consider extending these possibilities with two real scalar fields or a complex scalar field in FLRW cosmology, following the lines of the present work. The subject is of current interest, since cosmology with two or more fields has been recently studied under interesting new directions, in particular, in \cite{2F,2F1,PT,2F2} and in references therein. 
Some of those new possibilities and other related issues are presently under consideration, and we hope to report on them in the near future.

\section*{Acknowledgments}

We would like to thank CNPq, FAPESP, PADCT/CNPq and PRONEX/CNPq/FAPESQ for financial support. ABP thanks E. Abdalla for helping him to visit
Departamento de F\'\i sica, Universidade Federal da Para\'\i ba.

 
\end{document}